\documentclass[a4paper,12pt]{article}

\usepackage{amsthm,amsmath,amssymb,amsfonts}

\usepackage{graphicx}
\usepackage{caption}
\usepackage{subcaption}

\providecommand*{\pderiv}[3][]{%
        \frac{\partial^{#1}{#2}}%
                {\partial {#3}^{#1}}}

\def \ring {{\cal R}}

\def\bbbc{{\mathbb C}}
\def\bbbz{{\mathbb Z}}

\def\hu{{\hat{u}}}
\def\dd{D_x}
\def\bu{{\bf m}}
\def\bbu{{U}}
\def\bbv{{V}}
\def\du{{m}}
\def\dv{{n}}
\def\pu{{u}}
\def\pv{{v}}
\def\hu{{\hat{u}}}
\def\hv{{\hat{v}}}

\def\cH{{\cal H}}

\def\cJ{{\cal J}}
\def\s3{\sqrt{3}}
\def\s5{\sqrt{5}}

\newtheorem{Def}{Definition}
\newtheorem{The}{Theorem}
\newtheorem{Pro}{Proposition}

\newtheorem{Rem}{Remark}
\newtheorem{Cor}{Corollary}

\newcommand{\bear}{\begin{array}}
\newcommand{\eear}{\end{array}}

\newenvironment{prf}{\trivlist \item [\hskip
\labelsep {\bf Proof:}]\ignorespaces}{\qed \endtrivlist}

\newfont{\tenbi}{cmbxti10}
\newcommand{\beq}{\begin{equation}}
\newcommand{\eeq}{\end{equation}}

\newcommand{\rC}{\mathrm{C}}

\newcommand{\rd}{\mathrm{d}}

\newcommand\la{{\lambda}}

\newcommand\ka{{\kappa}}
\newcommand\al{{\alpha}}
\newcommand\be{{\beta}}

\newcommand\om{{\omega}}

\newcommand\si{{\sigma}}

\newcommand\ups{{\vartheta}}

\begin{document}
\title{Two-component generalizations of the Camassa-Holm equation}
\author{Andrew N.W. Hone\thanks{School of Mathematics, 
Statistics \& Actuarial Science, University of Kent, Canterbury, CT2 7NF, UK.}, 
Vladimir Novikov\thanks{Department of Mathematical Sciences, Loughborough University, 
Loughborough, LE11 3TU, UK.} and Jing Ping Wang$^*$
       }

\maketitle
\begin{abstract} 
A classification of integrable two-component systems of non-evolutionary partial differential equations that are 
analogous to the Camassa-Holm equation is carried out via the perturbative symmetry approach. 
Independently, a classification of compatible pairs of Hamiltonian operators is carried out, which leads to bi-Hamiltonian structures for the same systems of equations. Some exact solutions  and  Lax pairs are also constructed 
for the systems considered.
\end{abstract} 

\section{Introduction}

In recent years there has been a growing interest in integrable
non-evolutionary partial differential equations of  the form
\begin{equation}
\label{CHt} (1-D_x^2)u_t=F(u,u_x,u_{xx},u_{xxx},\ldots), \quad
u=u(x,t),\quad D_x=\frac{\partial}{\partial x},
\end{equation}
where $F$ is some function of $u$ and its derivatives with respect to
$x$. The most celebrated example of this type of equation is the
Camassa--Holm equation \cite{CH}:
\beq\label{caho}
(1-D_x^2)u_t=3uu_x-2u_xu_{xx}-uu_{xxx}.
\eeq
Other examples of integrable equations of the form (\ref{CHt}) include the Degasperis-Procesi equation 
\[
(1-D_x^2)u_t=4uu_x-3u_xu_{xx}-uu_{xxx},
\]
(see \cite{dp, DHH})
as well as equations with cubic nonlinearity, such as 
\begin{eqnarray*}
\label{A}
(1-D_x^2)u_t&=&u^2u_{xxx}+3uu_xu_{xx}-4u^2u_x,\\ \nonumber \\
\label{B}
(1-\epsilon^2D_x^2)u_t&=&D_x\left(u^2u_{xx}-u_x^2u_{xx}+uu_x^2-u^3\right)
\end{eqnarray*}
(see  \cite{HW, Novikov} and \cite{fokas, Qiao},  respectively).
All the of the latter equations of Camassa-Holm type are integrable by the inverse scattering transform. 
They possess infinite hierarchies of local conservation laws and (quasi-)local higher symmetries, bi-Hamiltonian structures and other remarkable attributes of integrable systems. Part of the fascination with these sorts of equations is due to the fact 
that as well as having traditional (smooth) multi-soliton solutions, they admit weak solutions of peakon (peaked soliton) type, and also display interesting blowup and wave-breaking phenomena \cite{mckean}. The complete classification of integrable equations of the form (\ref{CHt}) was carried out in \cite{Novikov} using the perturbative symmetry approach introduced in \cite{MN}. 
Various approaches to generating multicomponent systems of Camassa-Holm type have been proposed recently, based on 
energy-dependent spectral problems \cite{HI}, or Novikov algebras \cite{strachan}. 

In this paper we study integrable two-component systems of the form
\begin{equation}
\label{CHsys} \left\{
\begin{array}{c}
(1-D_x)u_t=F(u,v,u_x,v_x,u_{xx},v_{xx}),\\
(1+D_x)v_t=G(u,v,u_x,v_x,u_{xx},v_{xx}),
\end{array}\right.
\end{equation}
where $F,G$ are polynomials over $\bbbc$ in $u,v$ and their $x$-derivatives. An example of an integrable system of the form (\ref{CHsys}) is
\begin{equation}
\label{CLZ} \left\{
\begin{array}{c}
(1-D_x)u_t=2(u+v)u_x-(u+v)u_{xx}-u_x^2,\\
(1+D_x)v_t=2(u+v)v_x+(u+v)v_{xx}+v_x^2.
\end{array}
\right.
\end{equation}
The above system 
is related
 to a system which  (up  to sending $t\to -t$ and renaming variables) was given as 
\begin{equation}
\label{CLZ1}
\left\{\begin{array}{rcl} m_t & =& pm_x+2mp_x-qq_x, \\
q_t& =&(pq)_x, \qquad  m=(1-D_x^2) p,
\end{array}
\right.
\end{equation}
by Chen, Liu and Zhang \cite{CLZ},   
and related to an alternative 
system of the form (\ref{CHsys}) presented by Falqui \cite{Falq}, namely 
\beq\label{falqui} 
 \left\{
\begin{array}{rcl}
(1-D_x)U_t&=&V_x+2UU_x-UU_{xx}-U_x^2,\\
(1+D_x)V_t&=&2U_xV+2UV_x+UV_{xx}+U_xV_x 
\end{array}
\right.
\eeq
(again, up to renaming variables, and fixing the value of a 
parameter).
To be precise, under the  transformation
\beq\label{1miu}
p=u+v,\qquad q=(1-D_x)u+(1+D_x)v, 
\eeq
which is of Miura type, solutions of the system  (\ref{CLZ}) are mapped to solutions of  (\ref{CLZ1}), while 
$$ 
p=U, \qquad q^2=\Big((1-D_x)U\Big)^2 -2(1+D_x)V
$$ 
is a Miura map from (\ref{falqui})  to  (\ref{CLZ1}).


The rest of the paper is concerned with 
classifying integrable systems of the form (\ref{CHsys}). 
In the next section we outline the perturbative symmetry approach in the context of non-evolutionary systems with two dependent variables, 
and explain how it leads to an integrability test for such systems. 
Section 3 contains the result of applying this integrability test, in the form of a list of systems with quadratic, cubic and mixed quadratic/cubic 
nonlinear terms; there are six systems in total, presented in Theorems 2,3 and 4 below. The fourth section is concerned with a different problem, namely that of classifying pairs of compatible Hamiltonian operators in 
two dependent variables. However, this turns out to be highly relevant to the preceding considerations, since it provides a bi-Hamiltonian structure for (almost) 
every system in the aforementioned list. In the fifth section we consider changes of independent variables, 
specifically reciprocal transformations (sending conservation laws to conservation laws); these are helpful for the construction  
of Lax pairs and exact solutions,  which we illustrate in some cases. The paper ends with  conclusions and suggestions for future work.

\section{Integrability test: perturbative symmetries}

In this section we briefly recall the basic definitions and notations of the perturbative symmetry approach (for details see \cite{MN, review}). 
We also present the integrability test 
which we will subsequently apply to isolate integrable generalizations of  the Camassa--Holm equation.

\subsection{Quasi-local polynomials and definition of symmetries}
Let $u, v$ be functions in $x,t$. Polynomials in $u, v$ and their $x$-derivatives over $\bbbc$
form a differential ring $\ring$ with an $x$-derivation
$$D_x=\sum_{k=0}^\infty (u_{k+1}  \pderiv{}{u_k}+v_{k+1}  \pderiv{}{v_k}),  $$  
where $u_k, v_k$ denote $k$-th derivatives of $u, v$ with respect to $x$.
In particular, $u_0$ and $v_0$ denote the functions $u$ and $v$ themselves.
We often omit the zero index of $u_0$ and $v_0$ and simply write $u$ and $v$.

We will assume that $1\not\in
\ring$. Elements of the ring $\ring$ are finite sums of monomials in $u, v$ and their $x$-derivatives
with complex coefficients. The degree of a monomial
is defined as a total power, i.e. the sum of all powers of 
variables that contribute to the monomial. Let ${\cal R}^n$ denote
the set of polynomials of degree $n$ in $u,v$ and their $x$-derivatives. Then ring $\ring$
has a gradation 
\[ \ring = \bigoplus_{n\in \bbbz_+}{\cal R}^n \, ,
\quad {\cal R}^n \cdot {\cal R}^m\subset {\cal R}^{n+m} \, .
\]
 Elements of ${\cal R}^1$ are linear
functions of the $u,v$ and their derivatives, elements of ${\cal R}^2$ are quadratic, etc.
It is convenient to define a
``little-oh'' order symbol $o({\cal R}^n)$. We say that $f=o({\cal
R}^n)$ if $f\in \bigoplus_{k>n}{\cal R}^k$, i.e. the degree of every
monomial of $f$ is bigger than $n$.

Since $1\not\in\ring$, the kernel of the linear map
$D_x:\ring\mapsto\mbox{Im}\, D_x\subset\ring$ is empty and
therefore $D_x^{-1}$ is defined uniquely on $\mbox{Im}\, D_x$.

To an element $g\in\ring$ we associate  differential operators $g_{*,u}$ and $g_{*,v}$ called
Fr\'echet derivatives with respect to $u$ and $v$ and  defined as
\[
g_{*,u}=\sum_{k\ge 0}\frac{\partial g}{\partial u_k}D_x^k,\qquad
g_{*,v}=\sum_{k\ge 0}\frac{\partial g}{\partial v_k}D_x^k.
\]

Now we need to introduce a concept of quasi-local differential polynomials and the corresponding extension of the ring $\ring$. The idea of this extension is similar to that in  \cite{MY,wang06,MN}.

To rewrite the Camassa-Holm type system (\ref{CHsys}) in evolutionary form, we
introduce a pair of pseudo-differential operators
\begin{eqnarray}\label{delta}
 \Delta_{-}=(1-D_x)^{-1},\quad   \Delta_{+}=(1+D_x)^{-1}.
\end{eqnarray}
System (\ref{CHsys}) then can be  rewritten as
\begin{eqnarray}\label{cahs}
 \left\{ \begin{array}{l} u_t=\Delta_{-} F(u,v, u_1,v_1, \cdots, u_n,
v_m)\\v_t=\Delta_{+} G(u,v, u_1,v_1, \cdots, u_n, v_m)
\end{array}\right. .
\end{eqnarray}
Clearly, if $F,G\in\ring$ then the right hand side of the system (\ref{cahs}) no longer consists of differential polynomials and we need an extension of the original differential ring $\ring$.

Consider  the following sequence of ring extensions:
\[
\ring_{0}=\ring,\quad \ring_{1}=\overline{\ring_{0}\bigcup
\Delta_{+}(\ring_{0})\bigcup
\Delta_{-}(\ring_{0})},\quad
\ring_{n+1}=\overline{\ring_{n}\bigcup
\Delta_{+}(\ring_{n})\bigcup
\Delta_{-}(\ring_{n})},
\]
where the set $\Delta_{\pm}(\ring_{n})=\{\Delta_{\pm}(a): a\in \ring_{n}\}$
and
the horizontal line denotes the ring closure.
The index $n$ indicates the ``nesting depth" of operators
$\Delta_{\pm}$. We then define quasi-local differential polynomials as follows. 
\begin{Def} An element $f$ is called a quasi-local differential polynomial if $f\in\ring_{n}$ for sufficiently large $n$.
\end{Def}

The right-hand side of equations in (\ref{cahs}) lies in $\ring_{1}$.  Its
symmetries and densities of conservation are also generally speaking all quasi-local and 
 belong to $\ring_{k}$ for
some $k\ge 0$ .

We now recall the  definition
of a symmetry.
\begin{Def}\label{sym}
A pair of quasi-local differential polynomials $P$ and $Q$
is called a symmetry of an evolutionary system $u_t=f, v_t=g$, where $f,g$ are quasi-local polynomials, if the system
\[
u_{\tau}=P, \ v_{\tau}=Q
\]
is compatible with $u_t=f,v_t=g$.
\end{Def}

If ${\bf a}=
(f,g)^T$ and ${\bf
b}=(P,Q)^T$ then the above definition is equivalent to the vanishing of the Lie bracket
\begin{eqnarray}\label{lieD}
 [{\bf a},{\bf b}]=\left(\begin{array}{cc} f_{*,u} & f_{*,v}\\ g_{*,u} &
g_{*,v}\end{array}\right)\left(\begin{array}{l} P\\ Q \end{array}\right)
-\left(\begin{array}{cc} P_{*,u} & P_{*,v}\\ Q_{*,u} & Q_{*,v}\end{array}
\right)\left(\begin{array}{l} f\\ g \end{array}\right)\ .
\end{eqnarray}

We finally define a notion of formal pseudodifferential series (or just formal series) as an object of the form
\begin{equation}\label{fser}
A=\sum_{k\ge 0} a_{N-k} D_x^{N-k}\, ,
\end{equation}
with coefficients being quasi-local differential polynomials or constants. The order of the formal series (\ref{fser}) is $N$ (we assume that the leading
coefficient $a_N\neq 0$). The formal series form a ring:
the sum of formal series is
defined in the obvious way, while
 multiplication (composition) is defined by
\begin{equation}\label{fscomp}
 a_n D_x^n \circ b_m D_x^m=\sum _{k\ge
0}\left(\begin{array}{c}n\\k\end{array}\right)
a_n D_x^k (b_m)D_x^{m+n-k} \, .
\end{equation}
For positive $n$ the sum (\ref{fscomp}) is finite since the binomial
coefficients
\[  \left(\begin{array}{c}n\\k\end{array}\right)=
\frac{n(n-1)(n-2)\cdots (n-k+1)}{k!} \]
vanish for $k>n$, and for negative $n$ the composition is well-defined in the sense
of formal series.

In the symmetry approach we admit the following definition of integrability:

\begin{Def} System (\ref{cahs}) is {\it integrable} if it possesses
an infinite hierarchy of symmetries.
\end{Def}

In the following subsections we present the necessary conditions for existence of a hierarchy of symmetries. For this it is convenient to introduce the symbolic representation of the ring of quasi-local polynomials and derive the necessary conditions in the symbolic representation.

\subsection{Symbolic representation}

In this subsection we introduce the symbolic representation of the ring of differential polynomials and its extensions.
We first recall the symbolic representation $\hat{\cal R}$ of the ring $\ring$.
A symbolic representation
of a monomial $$u_0^{n_0}u_1^{n_1}\cdots u_p^{n_p}v_0^{m_0}v_1^{m_1}\cdots v_q^{m_q},\qquad n_0+n_1+\cdots
+n_p=n,\, m_0+m_1+\cdots+m_q=m$$ is defined as:
\begin{eqnarray}\nonumber
 &&u_0^{n_0}u_1^{n_1}\cdots
u_p^{n_p}v_0^{m_0}v_1^{m_1}\cdots v_q^{m_q}\to\\ \label{mon} &&\to
\hu^n \hv^m
\langle\xi_1^0\xi_2^0\cdots\xi_{n_0}^0\xi_{n_0+1}^1\cdots
\xi_{n_0+n_1}^1\cdots\xi_n^p\rangle_{\xi}\langle\zeta_1^0\zeta_2^0\cdots
\zeta_{m_0}^0\zeta_{m_0+1}^1
\cdots\zeta_{m_0+m_1}^1\cdots\zeta_m^q\rangle_{\zeta}\ ,
\end{eqnarray}
where triangular brackets $\langle\rangle_{\xi}$ and
$\langle\rangle_{\zeta}$ denote the averaging over the group $\Sigma_n$
of permutations of $n$ elements $\xi_1,\ldots,\xi_n,$ 
and the group $\Sigma_m$ of $m$ elements $\zeta_1,\ldots,\zeta_m$
respectively. That is $\langle c(\xi_1,\ldots,\xi_n,\zeta_1,\ldots
,\zeta_m)\rangle_{\xi}$ is
\[
\langle c(\xi_1,\ldots,\xi_n,\zeta_1,\ldots
,\zeta_m)\rangle_{\xi}=\frac{1}{n!}\sum_{\sigma\in\Sigma_n}c(\sigma(\xi_1),\ldots,\sigma(\xi_n),\zeta_1,\ldots
,\zeta_m)
\]
and the similar definition holds for averaging with respect to $\zeta$ arguments. 
Later we refer to this as symmetrisation operation.
For example, linear monomials $u_n, v_m$ are represented by
\begin{equation}\label{unvm}
u_n\to \hu\xi_1^n,\quad v_m\to \hv\zeta_1^m
\end{equation}
and quadratic monomials $u_nu_m$, $u_nv_m$, $v_nv_m$  have the following symbols
\begin{equation}\label{monom}
u_nu_m\to\frac{\hu^2}{2}(\xi_1^n\xi_2^m+\xi_1^m\xi_2^n),\quad u_nv_m\to \hu \hv(\xi_1^n\zeta_1^m),
\quad
v_nv_m\to \frac{\hv^2}{2} (\zeta_1^n\zeta_2^m+\zeta_1^m\zeta_2^n)\ .
\end{equation}
To the sum of two elements of the ring corresponds  the sum of their
symbols. To the product of two elements $f,g\in\ring$ with symbols
$f\to \hu^n \hv^ma(\xi_1,\ldots,\xi_n,\zeta_1,\ldots,\zeta_m)$ and $g\to
\hu^p \hv^qb(\xi_1,\ldots,\xi_p,\zeta_1,\ldots,\zeta_q)$ corresponds
\begin{equation}\label{multmonoms}
fg\to \hu^{n+p} \hv^{m+q}\langle\langle
a(\xi_1,\ldots,\xi_n,\zeta_1,\ldots,\zeta_m)b(\xi_{n+1},\ldots,\xi_{n+p},
\zeta_{m+1},\ldots,\zeta_{m+q})\rangle_{\xi}\rangle_{\zeta},
\end{equation} where the symmetrisation operation is taken with respect to
permutations of all arguments $\xi$ and $\zeta$. It is easy to see
that the symbolic representations of quadratic (\ref{monom}) and general
(\ref{mon}) monomials immediately follow from (\ref{unvm}) and
(\ref{multmonoms}).

If $f\in\ring$ has a symbol $f\to
\hu^n \hv^ma(\xi_1,\ldots,\xi_n,\zeta_1,\ldots,\zeta_m)$, then the
symbolic representation for its $N$th derivative $D_x^N(f)$ is 
\[
D_x^N(f)\to
\hu^n \hv^m(\xi_1+\xi_2+\cdots+\xi_n+\zeta_1+\zeta_2+\cdots\zeta_m)^N
a(\xi_1,\ldots,\xi_n,\zeta_1,\ldots,\zeta_m) .
\]
We will assign a symbol $\eta$ to the operator $D_x$ in the symbolic
representation, with the 
action 
\[
\eta^N\cdot \hu^n \hv^ma(\xi_1,\ldots,\xi_n,\zeta_1,\ldots,\zeta_m)=\hu^n \hv^m
(\xi_1+\xi_2+\cdots+\xi_n+\zeta_1+\zeta_2+\cdots\zeta_m)^N
a(\xi_1,\ldots,\xi_n,\zeta_1,\ldots,\zeta_m)
\]
If $f\in\ring$ and $f\to \hu^n \hv^m
a_{n,m}(\xi_1,\ldots,\xi_n,\zeta_1,\ldots,\zeta_m)$ then for the
symbol of its Fr{\'e}chet derivatives $f_{*,u}$ and $f_{*,v}$ we
have
\[
f_{*,u}\to
n \hu^{n-1} \hv^ma_{n,m}(\xi_1,\ldots,\xi_{n-1},\eta,\zeta_1,\ldots,\zeta_m),\,
f_{*,v}\to
m\hu^n \hv^{m-1}a_{n,m}(\xi_1,\ldots,\xi_n,\zeta_1,\ldots,\zeta_{m-1},\eta).
\]
Thus we have described 
the symbolic representation $\hat{\cal R}$ of the
differential ring $\ring$. 

To construct the symbolic representation of the quasi-local
rings $\ring_k,\,k=1,2,\ldots$ it is enough to
note that the symbolic representation of operator $\Delta_{\pm}=(1\pm D_x)^{-1}$ is
\[
\Delta_{\pm}\to (1\pm\eta)^{-1}.
\]
Now if $f\in\ring$ and $f\to
\hu^n \hv^ma(\xi_1,\ldots,\xi_n,\zeta_1,\ldots,\zeta_m)$, then
\[
\Delta_{\pm}(f)\to
\hu^n \hv^m\frac{a(\xi_1,\ldots,\xi_n,\zeta_1,\ldots,\zeta_m)}{1\pm(\xi_1+\cdots+\xi_n+\zeta_1+\cdots+\zeta_m)} .
\]
Using  the addition and multiplication operations where  necessary, we
thus construct the symbolic representation of $\hat{\ring}_k,\,k=1,2,\ldots$.

Finally, we define the symbolic representation for pseudo-differential formal series.
For any two terms $fD_x^p,\,gD_x^q$ of formal series ($p,q\in{\bbbz}$ and $f,g\in \ring_k$) with symbols  
$$f\to \hu^n \hv^ma(\xi_1,\ldots,\xi_n,\zeta_1,\ldots,\zeta_m),\qquad g\to
\hu^s \hv^rb(\xi_1,\ldots,\xi_s,\zeta_1,\ldots,\zeta_r)$$
the composition rule in the symbolic representation reads
\begin{eqnarray} \nonumber
&&
\quad fD_x^p\circ gD_x^q \to \hu^n \hv^ma(\xi_1,\ldots,\xi_n,\zeta_1,\ldots,\zeta_m) \eta^p\circ
\hu^s \hv^rb(\xi_1,\ldots,\xi_s,\zeta_1,\ldots,\zeta_r) \eta^q\\ \nonumber
&& = 
\hu^{n+s} \hv^{m+r}\langle\langle
a(\xi_1,\ldots,\xi_n,\zeta_1,\ldots,\zeta_m)(\xi_{n+1}+ 
\cdots+\xi_{n+s}+\zeta_{m+1}+ 
\cdots+\zeta_{m+r}+\eta)^p\times\\  \label{multip} 
&&\times
\, b(\xi_{n+1},
\ldots,\xi_{n+s},\zeta_{m+1},
\ldots,\zeta_{m+r})\eta^q\rangle_{\xi}\rangle_{\zeta},
\end{eqnarray}  
where the symmetrisation is taken with respect to permutations of
arguments $\xi$ and arguments $\zeta$, but not the argument $\eta$.

More generally we consider formal series of the form
\begin{equation}\label{Aa}
A=a_{00}(\eta)+\hu a_{10}(\xi_1,\eta)+\hv a_{01}(\zeta_1,\eta)+\hu^2 a_{20}(\xi_1,\xi_2,\eta)+
\hu \hv a_{11}(\xi_1,\zeta_1,\eta)+\hv^2 a_{02}(\zeta_1,\zeta_2,\eta)+\cdots,
\end{equation}
where the coefficients
$a_{nm}(\xi_1,\ldots,\xi_n,\zeta_1,\ldots,\zeta_m,\eta)$ are formal series in $\eta$, i.e. 
\[
a_{nm}(\xi_1,\ldots,\xi_n,\zeta_1,\ldots,\zeta_m,\eta)=\sum_{k=k_{nm}}^{\infty}
a_{nm}^k(\xi_1,\ldots,\xi_n,\zeta_1,\ldots,\zeta_m)\eta^{-k},
\]
with $a_{nm}^k(\xi_1,\ldots,\xi_n,\zeta_1,\ldots,\zeta_m)$ being  
symmetric functions with respect to permutations of arguments
$\xi_i$ and arguments  $\zeta _i$.
Similar to the rule (\ref{multip}), the composition of two monomials is defined as
\[
\hu^n \hv^m a(\xi_1,\ldots,\xi_n,\zeta_1,\ldots,\zeta_m,\eta)\circ \hu^p \hv^q b(\xi_1,\ldots,\xi_p,\zeta_1,\ldots,\zeta_q,\eta) \]\[ = 
\hu^{n+p} \hv^{m+q}\langle\langle
a(\xi_1,\ldots,\xi_n,\zeta_1,\ldots,\zeta_m, \xi_{n+1}+\cdots+\xi_{n+p}+\zeta_{m+1}+
\cdots+\zeta_{m+q}+\eta ) \times \] \[ \times b(\xi_{n+1},\ldots,\xi_{n+p},
\zeta_{m+1},\ldots,\zeta_{m+q},\eta)\rangle_{\xi}\rangle_{\zeta}.
\]

\begin{Def}\label{localf}
We shall call a function
$a_{nm}(\xi_1,\ldots,\xi_n,\zeta_1,\ldots,\zeta_m,\eta)$ {\em quasi-local} if all
the coefficients
of its expansion
\begin{eqnarray}\label{local}
a_{nm}(\xi_1,\ldots,\xi_n,\zeta_1,\ldots,\zeta_m,\eta)=\sum_{k}
a_{nm}^k(\xi_1,\ldots,\xi_n,\zeta_1,\ldots,\zeta_m)\eta^{-k},\quad
\eta\to\infty
\end{eqnarray}
are the symbolic representations of some elements from $\ring_{k}$ for some $k\ge 0$.
\end{Def}

In particular, if all the coefficients in (\ref{local}) are symmetric polynomials in each of the two sets of variables
$\xi_1,\ldots,\xi_n$, and $\zeta_1,\ldots,\zeta_m$, we say that the function $a_{nm}(\xi_1,\ldots,\xi_n,\zeta_1,\ldots,\zeta_m,\eta)$
is local.

The set of formal series (\ref{Aa})  has the structure of an
associative noncommutative ring $\hat{\ring}_{\Delta} (\eta)$. It inherits the 
natural gradation from $\ring$, namely 
\[ \hat{\ring}_{\Delta}(\eta)=\bigoplus_{n=0}\hat{\ring}_{\Delta}^n(\eta), \]
where $\hat{\ring}_{\Delta}^n(\eta)$ with $n=0,1,2,3,\ldots$  are constant (i.e.
independent of $u,v$ ), linear in $u$ or $v$, quadratic, cubic,  etc. We say that a formal
series $A=o(\hat{\ring}_{\Delta}^n(\eta))$ if $A\in
\bigoplus_{k>n}\hat{\ring}_{\Delta}^k(\eta)$.

\subsection{Formal recursion operator and necessary  conditions for integrability}

In this subsection we formulate the necessary conditions for integrability of a system of the form  
\begin{eqnarray}\label{cahs1}
 \left\{ \begin{array}{l} u_t=\Delta_{-} f(u,v, u_1,v_1, \cdots, u_n,
v_m)\\v_t=\Delta_{+} g(u,v, u_1,v_1, \cdots, u_n, v_m)
\end{array}\right. .
\end{eqnarray}
Let $\hat{f},\,\hat{g}$ be the symbolic representations of the differential polynomials $f,g$, so that 
\begin{equation}
\label{rhs}
\begin{array}{c}
\hat{f}=\hu\omega_1(\xi_1)+\hv\omega_2(\zeta_1)+\hu^2a_{2,0}(\xi_1,\xi_2)+\hu\hv a_{1,1}(\xi_1,\zeta_1)+\hv^2a_{0,2}(\zeta_1,\zeta_2)+\cdots \\
\hat{g}=\hu\omega_3(\xi_1)+\hv\omega_4(\zeta_1)+\hu^2b_{2,0}(\xi_1,\xi_2)+\hu\hv b_{1,1}(\xi_1,\zeta_1)+\hv^2b_{0,2}(\zeta_1,\zeta_2)+\cdots 
\end{array}
\end{equation}
Define
$$
F=\left(\begin{array}{cc}\hat{f}_{*,u}&\hat{f}_{*,v}\\ \hat{g}_{*,u}&\hat{g}_{*,v}\end{array}\right)
$$
and let
\begin{equation}
\label{lambda}
\Lambda=\left(\begin{array}{cc}L^{(1)}&L^{(2)}\\ L^{(3)}&L^{(4)}\end{array}\right),
\end{equation}
where $L^{(i)},\,i=1,\ldots,4$ are formal series, 
$$
L^{(i)}=\phi^{(i)}(\eta)+\hu\phi^{(i)}_{10}(\xi_1,\eta)+\hv\phi^{(i)}_{01}(\zeta_1,\eta)+\hu^2\phi^{(i)}_{20}(\xi_1,\xi_2,\eta))+\hu\hv\phi^{(i)}_{11}(\xi_1,\zeta_1,\eta)+\hv^2\phi^{(i)}_{02}(\zeta_1,\zeta_2,\eta)+\cdots.
$$
\begin{Def} A formal series $\Lambda$ (\ref{lambda}) is called a formal recursion operator  for system (\ref{cahs1}) if all the coefficients $\phi^{(i)}_{jk}$ are quasi-local and it satisfies the equation
\begin{equation}
\label{eqlambda}
\Lambda_t=F\circ\Lambda-\Lambda\circ F
\end{equation}
\end{Def}
In the above definition $\Lambda_t$ stands for a formal series obtained from $\Lambda$ by differentiating all the coefficients $\phi_{jk}^{(i)}$ by $t$ and replacing $u_t$ and $v_t$ according  to the system (\ref{cahs1}).

\begin{The}
Assume that the system (\ref{cahs1}) is such that
\begin{equation}
\label{condomega}
\omega_2(\zeta)=0 =\omega_3(\xi)=0
\end{equation}
and $\omega_1(\xi)\ne c_1\xi,\,\,\omega_4(\zeta)\ne c_4\zeta$ (for constants $c_1,c_4$).
Suppose that the system (\ref{cahs1}) possesses an infinite hierarchy of quasi-local higher symmetries. Then the system possesses a formal recursion operator (\ref{lambda}) with $\phi^{(2)}(\eta)=\phi^{(3)}(\eta)=0$ and $\phi^{(1)}(\eta)=\phi^{(4)}(\eta)=\eta$. 
\end{The}

The assumption (\ref{condomega}) implies that the linear part of the system (\ref{cahs1}) is {\it diagonal}; in principle, 
this condition  may be removed. (Note that Falqui's system 
 (\ref{falqui}) is excluded by this assumption.) In the diagonal case the proof of the theorem is essentially the same as the proof of the analogous Theorem 2 from \cite{MN} and therefore we omit it here.

Theorem 1 provides the necessary integrability conditions for the system (\ref{cahs1}). These can be obtained as follows:
\begin{itemize}
\item For a given system (\ref{cahs1}) one solves the equation (\ref{eqlambda}) with respect to $\Lambda$ and finds $\phi^{(i)}_{jk}(\xi_1,\ldots,\xi_j,\zeta_1,\zeta_k,\eta)$.
\item One then verifies the quasi-locality conditions of $\phi^{(i)}_{jk}(\xi_1,\ldots,\xi_j,\zeta_1,\zeta_k,\eta)$ and obtains the obstructions to integrability (if any) for the system (\ref{cahs1}).
\end{itemize}

To classify integrable systems of the Camassa-Holm type (see the next section) we only need to verify quasi-locality of $\phi^{(i)}_{jk},\,\,i=1,\ldots,4$ with $j+k\le 3$.

\section{Classification theorems}\label{sec3}

In this section we present the classification of integrable Camassa-Holm type systems of the form
\begin{equation}
\label{chgensys}
\left\{\begin{array}{c}
(1-D_x)u_t=\lambda_1 u_1+\lambda_2 u_2+f\\
(1+D_x)v_t=\mu_1 v_1+\mu_2 v_2+g
\end{array}\right. ,
\end{equation}
where $f,g$ are polynomials containing terms of degree two or above in $u,v,u_1,v_1,u_2,v_2$. We will also assume that $\lambda_2\ne -\lambda_1$ and $\mu_2\ne \mu_1$ as otherwise the linear part of each equation of the system will be $\lambda_1(1-D_x)u_1$ and $\mu_1(1+D_x)v_1$, and individually these terms are removable by a Galilean transformation.


We will restrict the classification to non-linearisable systems and therefore require the existence of non-trivial conservation laws. This allows us to further restrict the admissible linear terms in (\ref{chgensys}). 

\begin{Pro}
If the system (\ref{chgensys}) possesses a 
conservation law with nonlinear density $\rho$ 
then $\mu_2=-\lambda_2$ and $\mu_1=\lambda_1$.
\end{Pro}

\begin{prf} 
 Rewriting the system (\ref{chgensys}) in evolutionary form and transforming the system to the symbolic representation we obtain
\begin{equation}
\label{chgensyssym}
\left\{\begin{array}{c}
\hu_t=\hu \omega_1(\xi_1)+\hat{f}\\
\hv_t=\hv \omega_2(\zeta_1)+\hat{g}
\end{array}\right. ,
\end{equation}
where
$$
\omega_1(k)=\frac{\lambda_1 k+\lambda_2 k^2}{1-k},\quad \omega_2(k)=\frac{\mu_1 k+\mu_2 k^2}{1+k}
$$
and $\hat{f},\,\hat{g}$ are the symbolic representations of the $f$ and $g$. Clearly, the condition $\lambda_2\ne -\lambda_1$ and $\mu_2\ne \mu_1$ implies that 
$\omega_{1,2}(k)\ne c_{1,2}k$ for  constants $c_{1,2}$.
Assume first that $\rho$ is a density of a conservation law with symbol
$
\hat{\rho}=\hu^n\hv^ma(\xi_1,\ldots,\xi_n,\zeta_1,\ldots,\zeta_m)+o(\hat{\ring}^{n+m}).
$
Then we must have $\rho_t\in \mbox{Im}(D_x)$. In the symbolic representation we have
$$
\hat{\rho}_t=\hu^n\hv^m a(\xi_1,\ldots,\xi_n,\zeta_1,\ldots,\zeta_m)\left[\omega_1(\xi_1)+\cdots+\omega_1(\xi_n)+\omega_2(\zeta_1)+\cdots+\omega_2(\zeta_m)\right] 
+o(\hat{\ring}^{n+m}).
$$
Since $\rho_t\in \mbox{Im}(D_x)$ we must have
$$
(\xi_1+\cdots+\xi_n+\zeta_1+\cdots+\zeta_m)\,|\,a(\xi_1,\ldots,\xi_n,\zeta_1,\ldots,\zeta_m)\left[\omega_1(\xi_1)+\cdots+\omega_1(\xi_n)+\omega_2(\zeta_1)+\cdots+\omega_2(\zeta_m)\right]
$$
Since $\rho$ is a non-trivial density, $\rho\notin\mbox{Im}(D_x)$ so that 
$a(\xi_1,\ldots,\xi_n,\zeta_1,\ldots,\zeta_m)$ is not divisible by $(\xi_1+\cdots+\xi_n+\zeta_1+\cdots+\zeta_m)$, 
and therefore we must have
$$
(\xi_1+\cdots+\xi_n+\zeta_1+\cdots+\zeta_m)\,|\,\omega_1(\xi_1)+\cdots+\omega_1(\xi_n)+\omega_2(\zeta_1)+\cdots+\omega_2(\zeta_m).
$$
This implies
$$
\omega_1(\xi_1)+\cdots+\omega_1(\xi_n)+\omega_2(\zeta_1)+\cdots+\omega_2(\zeta_{m-1})+\omega_2(-\xi_1-\cdots-\xi_n-\zeta_1-\cdots-\zeta_{m-1})=0.
$$
Assume that  $n+m>2$. Then differentiating the above expression  with respect to any two distinct arguments 
in succession we obtain either $\omega_1''=0$ or $\omega_2''=0$, which contradicts the conditions 
$\omega_{1,2}(k)\ne  c_{1,2}k$. Clearly, if $\rho$ is a density of a non-trivial conservation law with  symbol 
$$
\hat{\rho}=\sum_{i=0}^n\hu^i\hv^{n-i}a_i(\xi_1,\ldots,\xi_i,\zeta_1,\ldots,\zeta_{n-i})+o(\hat{\ring^{n}})
$$
and $n>2$ then we again arrive at the same contradiction.
So  the density must necessarily start with quadratic terms, and therefore
$
\hat{\rho}=\hu^2\,a(\xi_1,\xi_2)+\hu\hv\, b(\xi_1,\zeta_1)+\hv^2\,c(\zeta_1,\zeta_2)+o(\hat{\ring^2})$.
Hence
$$
\hat{\rho}_t=\hu^2\,a(\xi_1,\xi_2)[\omega_1(\xi_1)+\omega_1(\xi_2)]+\hu\hv\, b(\xi_1,\zeta_1)[\omega_1(\xi_1)+\omega_2(\zeta_1)]+\hv^2\,c(\zeta_1,\zeta_2)[\omega_2(\zeta_1)+\omega_2(\zeta_2)]+o(\hat{\ring^2}).
$$
Since $\rho_t\in\mbox{Im}{D_x}$ we must have
$$
a(\xi_1,-\xi_1)[\omega_1(\xi_1)+\omega_1(-\xi_1)]=0,\quad c(\zeta_1,-\zeta_1)[\omega_2(\zeta_1)+\omega_2(-\zeta_1)]=0,
$$
and
$b(\xi_1,-\xi_1)[\omega_1(\xi_1)+\omega_2(-\xi_1)]=0$.
If $\omega_1(\xi_1)+\omega_1(-\xi_1)=0$ then $\lambda_2=-\lambda_1$, which contradicts the assumptions. Thus $a(\xi_1,-\xi_1)=0$ and we can disregard this term as trivial. Similarly, if $\omega_2(\zeta_1)+\omega_2(-\zeta_1)=0$ then $\mu_2=\mu_1$, and thus we have $c(\zeta_1,-\zeta_1)=0$ and disregard this term as well. Therefore we must have 
$b(\xi_1,-\xi_1)\ne 0$,
The last condition gives $
\mu_2=-\lambda_2$ and $\mu_1=\lambda_1$. \end{prf} 

Using the above proposition with a combination of a Galilean transformation and rescaling $t$, 
we thus can consider only systems of the form
\begin{equation}
\label{chgensysred}
\left\{\begin{array}{c}
(1-D_x)u_t=u_1+f,\\
(1+D_x)v_t=v_1+g,
\end{array}\right. 
\end{equation}
where we assume that $f,g$ are polynomials in $u,v,u_1,v_1,u_2,v_2$ with  only quadratic terms, cubic terms, or both, 
and we consider these different cases separately.

Applying the integrability test as described above  leads to a total of six different systems, which are listed below according to the 
type of nonlinearity.

\vspace{.5in}
\underline{Systems with quadratic nonlinearity:}

\begin{The}\label{th2}
If the system (\ref{chgensysred}) with $f,g$ being {\it quadratic} polynomials in $u,v,u_1,v_1,u_2,v_2$ possesses an infinite hierarchy of higher symmetries then modulo the scaling transformations $u\to\alpha u,\,v\to\beta v,\,x\to\gamma x,\,t\to\delta t$ it is one of the list
\begin{equation}
\label{sysq1}\left\{\begin{array}{l} (1-D_x)u_t=u_1+D_x(2-D_x)u^2+2vD_x(2-D_x)u,\\ 
(1+D_x)v_t=v_1+2uD_x(2+D_x)v+D_x(2+D_x)v^2; \end{array}\right.
\end{equation}
\begin{equation}
\label{sysq2}
\left\{\begin{array}{l} (1-D_x)u_t=u_1+D_x(2-D_x)u^2+2D_xv(2-D_x)u,\\ 
(1+D_x)v_t=v_1+2D_xu(2+D_x)v+D_x(2+D_x)v^2. \end{array}\right.
\end{equation}
\end{The}

\underline{Systems with cubic nonlinearity:}

\begin{The}\label{th3}
If the system (\ref{chgensysred}) with $f,g$ being {\it cubic} polynomials in $u,v,u_1,v_1,u_2,v_2$ possesses an infinite hierarchy of higher symmetries then modulo the scaling transformations $u\to\alpha u,\,v\to\beta v,\,x\to\gamma x,\,t\to\delta t$ it is one of the list
\begin{equation}
\label{sysc1} \left\{\begin{array}{l} (1-D_x)u_t=u_1+vD_x(2-D_x)u^2,\\
(1+D_x)v_t=v_1+uD_x(2+D_x)v^2;
\end{array}
\right.
\end{equation}
\begin{equation}
\label{sysc2} \left\{\begin{array}{l} (1-D_x)u_t=u_1+D_xv(2-D_x)u^2,\\
(1+D_x)v_t=v_1+D_xu(2+D_x)v^2.
\end{array}
\right.
\end{equation}
\end{The}

\underline{Mixed nonlinearity:}

\begin{The}\label{th4}
If the system (\ref{chgensysred}) with $f,g$ being {\it mixed quadratic and cubic} polynomials in $u,v,u_1,v_1,u_2,v_2$ possesses an infinite hierarchy of higher symmetries then modulo the scaling transformations $u\to\alpha u,\,v\to\beta v,\,x\to\gamma x,\,t\to\delta t$ it is one of the list
\begin{equation}
\label{sysm1} \left\{\begin{array}{l} (1-D_x)u_t=u_1+\alpha D_x(2-D_x)u^2+2\beta D_xv(2-D_x)u+\gamma D_xv(2-D_x)u^2,\\
(1+D_x)v_t=v_1+2\alpha D_xu(2+D_x)v+\beta D_x(2+D_x)v^2+\gamma D_xu(2+D_x)v^2;
\end{array}
\right.
\end{equation}
\begin{equation}
\label{sysm2} \left\{\begin{array}{l} (1-D_x)u_t=u_1+\alpha D_x(2-D_x)u^2+2\beta v D_x(2-D_x)u+\gamma v D_x(2-D_x)u^2,\\
(1+D_x)v_t=v_1+2\alpha u D_x(2+D_x)v+\beta D_x(2+D_x)v^2+\gamma u D_x(2+D_x)v^2.
\end{array}
\right.
\end{equation}
\end{The}

 In the next section we consider compatible pairs of Hamiltonian operators, which will lead to bi-Hamiltonian structures for each of the six systems listed above.

\section{Compatible Hamiltonian Operators}
In this section, we attempt to classify integrable two-component Camassa-Holm 
equations based on bi-Hamiltonian structures. This is similar to the approach of \cite{strachan}, based 
on Novikov algebras; however, in due course we will obtain systems with cubic nonlinearity, which do not appear 
in the latter approach. 
  
We use the multivector method, as 
described in the standard reference \cite{mr94g:58260}, to investigate the 
conditions such that 
the specified types of antisymmetric operators $\cH$ with entries $\cH_{ij}$, $i,j=1,2$, depending on a pair of fields $m,n$, 
form Hamiltonian pairs with a 
nondegenerate constant-coefficient differential Hamiltonian operator 
given by
\begin{eqnarray}\label{conH}
 \cJ=\left(\begin{array}{cc} c_1 \dd-c_2 \dd^3 & c_3 \dd -c_4 \dd^2 \\
 c_3 \dd +c_4 \dd^2 & c_5 \dd-c_6 \dd^3
\end{array} \right), \qquad \text{with} \,\, \text{constants} \,\, c_i, \, i=1,\ldots, 6.
\end{eqnarray}
For the purpose of deriving coupled two-component Camassa-Holm equations, we 
are going to study three cases: 
\begin{eqnarray}\label{cases}
({\rm i})\ c_4=1;\quad  ({\rm ii})\ c_4=0, c_2=1; \quad ({\rm iii})\ c_4=c_2=0, 
c_6=1.
\end{eqnarray}
Moreover, we also use  {\it elimination requirements} to get rid of 
non-coupled (triangular) or non-Camassa-Holm type equations by removing pairs satisfying  one or more of the 
conditions 
\begin{itemize}
 \item $c_1 c_2=c_3 c_4=c_5 c_6=0$;
 \item The determinant of $\cJ$ is a multiple of $\dd$;
 \item  $(\cH_{11})_{\star,n}=(\cH_{12})_{\star,n }=0$ and 
$\cH_{11}\cJ_{12}-\cH_{12}\cJ_{11}=0$;
\item $(\cH_{21})_{\star,m}=(\cH_{22})_{\star,m }=0$ and 
$\cH_{21}\cJ_{22}-\cH_{22}\cJ_{21}=0$. 
\end{itemize}
Otherwise, we refer to the  Hamiltonian pairs $\cH$ and $\cJ$ as non-trivial CH Hamiltonian 
pairs.


\subsection{Compatible linear Hamiltonian operators}\label{sec41}
We consider  linear antisymmetric differential operators in 
dependent variables $\du$, $\dv$, of the form
\begin{eqnarray}\label{HH}
 \cH=\left(\begin{array}{cc} a_1 (\du \dd+\dd\du)+a_2 (\dv \dd+\dd \dv) & 
a_3 \du \dd +a_4 \du_1+a_5 \dv \dd +a_6 \dv_1 \\
a_3 \dd \du -a_4 \du_1+a_5 \dd \dv -a_6 \dv_1 & a_7 (\du \dd+\dd \du)+a_8 (\dv 
\dd+\dd \dv)
\end{array} \right),
\end{eqnarray}
where $a_i, i=1,\ldots, 8$ are constants.

\begin{The}\label{thH1}
Let the operators $\cH$ and $\cJ$ be given by 
(\ref{HH}) and  
(\ref{conH}), respectively.
\begin{itemize}
 \item Suppose that $c_4= 1$. There are three non-trivial CH Hamiltonian pairs:
\begin{enumerate}
 \item[{\rm (i)}] $
           \cH^{(1)}\!=\!\left(\!\!\!\begin{array}{cc} a_1 (\du \dd+\dd \du) & 
a_1 \dv \dd  \\a_1 \dd \dv  &0 \end{array} \!\!\!\right),  
\ \cJ^{(1)}=\left(\!\!\!\begin{array}{cc} c_1 \dd-c_2\dd^3 
& c_3\dd - \dd^2 \\ c_3 \dd + \dd^2 & c_5 \dd
\end{array}\!\!\! \right), \ a_1 c_2\neq 0;
           $
\item[{\rm (ii)}] $
           \cH^{(2)}\!=\!\left(\!\!\!\begin{array}{cc} 0 & a_8 \dd \du\\ a_8 \du 
\dd & 
a_8 (\dv \dd+\dd \dv) 
\end{array} \!\!\!\right), \ \cJ^{(2)}\!=\!\left(\!\!\!\begin{array}{cc} c_1 
\dd &c_3\dd - \dd^2 \\ c_3 \dd + \dd^2  & c_5 \dd - c_6 \dd^3
\end{array} \!\!\!\right), \ a_8 c_6\neq 0;
           $
\item[{\rm (iii)}] $
           \cH^{(3)}\!\!=\!\!\left(\!\!\!\!\begin{array}{cc} a_1 (\du 
\dd+\dd \du) \!&\! a_1 \dv\dd + a_8 \dd \du\\ a_1 \dd \dv+a_8 \du \dd \!&\! a_8 
(\dv\dd+\dd \dv) 
\end{array} \!\!\!\right)\!, \  
\cJ^{(3)}\!\!=\!\!\left(\!\!\!\!\begin{array}{cc}  c_1 \dd \!&\! c_3 \dd - 
\dd^2 \\ c_3 \dd +\dd^2  \!&\! c_5 \dd
\end{array} \!\!\!\right)\!, 
           $  $a_1 a_8\neq 0$.
\end{enumerate}
\item  Suppose  that $c_4= 0$, $c_2=1$ and the parameter $c_3$ is arbitrary.
There are two  non-trivial CH 
Hamiltonian pairs:
\begin{enumerate}


\item[{\rm (iv)}] $\cH^{(4)}\!=\!\left(\!\!\begin{array}{cc} a_1 (\du\dd+\dd 
\du)+a_2(\dv \dd+\dd \dv) &  a_2 c_6(\du \dd+\dd \du)+a_6 (\dv\dd+\dd \dv)\\  
a_2 c_6(\du \dd+\dd \du)+a_6 (\dv\dd+\dd \dv) &  a_6 c_6 (\du\dd+\dd \du)+a_8 
(\dv \dd+\dd \dv) \end{array}\!\! \right),\\
\cJ^{(4)}\!\!=\!\!\left(\!\!\begin{array}{cc} c_1 
\dd\!- \!\dd^3 & 0  \\0  &\!\!\!\!\!\! c_1 c_6 \dd\!-\!c_6 \dd^3\end{array}
\right),\qquad  a_1 a_6-a_2^2 c_6+a_2 a_8-a_6^2=0\ \mbox{and} \  c_6\neq 0;$ 

\item[{\rm (v)}] $
           \cH^{(5)}\!=\!\left(\!\!\!\begin{array}{cc} a_1 (\du \dd+\dd \du) & 
a_1 \dv \dd  \\
a_1 \dd \dv  &0
\end{array} \!\!\!\right),  \ \cJ^{(5)}\!=\!\left(\!\!\!\begin{array}{cc} c_1 
\dd-\dd^3 
& c_3\dd  \\c_3 \dd & c_5 \dd
\end{array} \!\!\!\right), \ a_1 c_5\neq 0.
           $
 \end{enumerate}
 \item  Suppose  that $c_4=c_2=0$, $c_6=1$ and the parameter $c_3$ is  arbitrary. 
 There is only one non-trivial CH Hamiltonian pair:
\begin{enumerate}
 \item[{\rm (vi)}] $
\cH^{(6)}=\left(\!\!\!\begin{array}{cc}  0 & a_8\dd  \du  \\a_8  \du \dd & 
a_8 (\dv \dd+\dd \dv)\end{array} \!\!\!\right),  \
\cJ^{(6)}=\left(\!\!\!\begin{array}{cc} c_1 \dd & c_3 \dd  \\ c_3 \dd  & c_5 
\dd-\dd^3
\end{array} \!\!\!\right), \quad a_8 c_1\neq 0.   $
 \end{enumerate}
\end{itemize}
\end{The}
\begin{prf}
 For the operators (\ref{conH}) and  (\ref{HH}), acting on the univector 
$\bf \xi$, we have
 \begin{eqnarray}
 &&\cJ({\bf \xi})\!=\!\cJ\left(\!\!\!\begin{array}{c}\theta\\ 
\eta\end{array}\!\!\!\right)
\!=\!\left(\!\!\!\begin{array}{c}
 c_1 \theta_1-c_2 \theta_3 + c_3 \eta_1 -c_4 \eta_2   \\
  c_3 \theta_1 +c_4 \theta_2   + c_5 \eta_1-c_6 \eta_3  
\end{array}\!\!\!\right);\label{Jf}
\\
 && \cH({\bf \xi})\!=\!\cH\left(\!\!\!\begin{array}{c}\theta\\ 
\eta\end{array}\!\!\!\right)\!=\!\left(\!\!\!\begin{array}{c}P\\ 
Q\end{array}\!\!\!\right)\nonumber\\
&&
\!=\!\left(\!\!\!\begin{array}{c} 2 a_1 \du \theta_1+a_1 \du_1 \theta+2 a_2 \dv 
\theta_1+a_2 \dv_1 \theta+a_3 \du \eta_1 +a_4 \du_1\eta+a_5 \dv \eta_1 +a_6 
\dv_1\eta \\
a_3 \du \theta_1 \!+\!(a_3\!-\!a_4) \du_1\theta\!+\!a_5 \dv \theta_1\!+\!(a_5 
\!-\!a_6) \dv_1 
\theta \!+\!2 a_7 \du \eta_1\!+\!a_7 \du_1 \eta\!+\!2 a_8 \dv \eta_1\!+\!a_8 
\dv_1 \eta
\end{array} \!\!\!\right)\!\!,\label{Hf}
 \end{eqnarray}
where we used the notation $\dd^i \theta=\theta_i$ and $\dd^i \eta=\eta_i$.
We define the bivector associated to the operator $\cH$ by 
\begin{eqnarray*}
\Theta_{\cH}=\frac{1}{2} \int {\bf \xi}\wedge \cH({\bf \xi})
&=&
\!\!\!\int\!\!  (a_1 \du + a_2 \dv) \theta \wedge 
\theta_1\!+\!\Big((a_3-a_4) 
\du+(a_5-a_6)\dv\Big)\theta\wedge \eta_1\!
\nonumber\\
&&\!\! -\!(a_4 \du+a_6 \dv) \theta_1 \wedge 
\eta \!+\!(a_7 \du+a_8\dv) \eta \wedge \eta_1,\label{T2h}
\end{eqnarray*}
where $\int f=\int g$ denotes the equivalence relation $f\equiv g$ iff $f-g\in {\rm Im} \dd$.
The operator $\cH$ is Hamiltonian if and only if it satisfies the Jacobi 
identity, which is equivalent \cite{mr94g:58260} to  the
vanishing of the trivector 
\begin{eqnarray}
 &&
{\rm Pr}_{\cH({\bf \xi})} (\Theta_{\cH})\!\!=\!\!\!\int\!\! \{ (a_1 P + 
a_2 Q) \wedge\theta \wedge\theta_1\!+\!(a_3-a_4) 
P\wedge \theta\wedge \eta_1\!\}\nonumber \\
&&+\int\!\!\{(a_5-a_6)Q\wedge\theta\wedge \eta_1-\!(a_4 P+a_6 Q)\wedge \theta_1 
\wedge\eta \!+\!(a_7 P+a_8 Q)\wedge \eta \wedge \eta_1\},\label{tri}
\end{eqnarray}
where $P$ and $Q$ are given by (\ref{Hf}). We substitute them into it and 
simplify the expression, 
 which leads to an algebraic system for the constants $a_j, j=1, \ldots, 8$, that is 
\begin{eqnarray}\label{sys1}
 \left\{\!\!\begin{array}{l} 
a_1 a_5+2 a_2 a_8-2 a_2 a_3+2 a_2 a_4-a_5^2+a_5 a_6=0;\quad a_2 a_7-a_3 a_6+a_4 
a_6=0; \\ 
 2 a_2 a_7-a_1 a_3+2 a_1 a_4 -a_3 a_5+a_3 a_6=0;\   \ 2 a_1 a_7+a_3 a_8-a_3 
a_4-2 a_6 a_7=0;\\
a_1 a_6+a_2 a_8-a_5 a_6+a_6^2-a_2 
a_4=0; \quad 2 a_2 a_7+a_5 
a_8-a_4 a_5-2 a_6 a_8=0;
\\a_1 a_7+a_3 a_8-a_4 
a_8-a_3 a_4+a_4^2-a_5 a_7+a_6 a_7=0;\quad a_2 a_7-a_3 
a_6+a_4 a_6=0.
        \end{array}
 \right.
\end{eqnarray}
The system (\ref{sys1}) provides necessary and sufficient conditions for  
$\cH$ to be Hamiltonian. 
For the purposes of this theorem, we solve it together with the conditions for $\cH$  to be  
compatible with the constant Hamiltonian operator $\cJ$, that is, the 
trivector
${\rm Pr}_{\cJ({\bf \xi})} (\Theta_{\cH})$ vanishes \cite{mr94g:58260}. We 
carry out this calculation in the same way as for (\ref{tri})
and obtain 
an algebraic system for the constants $c_i, a_j$, for $i=1,\ldots, 6$, 
$j=1,\ldots, 8$, namely 
\begin{eqnarray}\label{sys2}
 \left\{\begin{array}{l}a_2 c_4=0;\ a_7 c_4 =0;\ a_6 c_4=0; \  a_1 c_3 + a_2 c_5
-a_3 c_1+a_4 c_1-a_5 c_3+a_6 c_3=0;\\
a_5 c_4-a_6 c_4-a_1 c_4=0;\ -a_2 c_6+a_3 c_2-a_4 c_2=0;\ a_3 c_2-2 a_4 c_2=0;\ a_3 c_4-a_4 c_4=0;\\
(a_8 -a_4)c_4 =0;\ -a_5 c_6+2 a_6 c_6=0;\ a_7 c_1+a_8 c_3-a_4  c_3 -a_6 c_5=0;\ -a_7 c_2+a_6 c_6=0.
\end{array}\right.
\end{eqnarray}
Upon solving the latter system together with  (\ref{sys1}), when
$c_4=1$ we get the following
three solutions after applying our elimination requirements:
\begin{enumerate}
 \item $a_5=a_1,\ a_2=a_3=a_4=a_6=a_7=a_8=c_6=0,\ a_1c_2\neq 0;$
 \item $a_3=a_4=a_8,\ a_1=a_2=a_5=a_6=a_7=c_2=0, \ a_8 c_6\neq 0;$
 \item $a_5=a_1,\ a_3=a_4=a_8,\ a_2=a_6=a_7=c_2=c_6=0,\ a_1 a_8\neq 0$;
\end{enumerate} 
these correspond to the three nontrivial CH Hamiltonian pairs  (i)-(iii) in the statement.
Similarly, we treat the other two cases with the help of the Maple package 
Gr{\"o}ebner and obtained the listed pairs (iv)-(vi). This completes the proof.
\end{prf}

Any compatible Hamiltonian pair $\cH$ and $\cJ$ which does not depend explicitly 
on the independent variables 
$x$ and 
$t$, with $\cJ$ nondegenerate,  leads to an 
integrable equation for the 
vector of dependent variables  $\bu$, that is 
\begin{eqnarray}\label{bi}
 \bu_t=\cH \cJ^{-1} (\bu_x).
\end{eqnarray}
In fact, for scalar $\bu$, the Camassa-Holm equation (\ref{caho}) was first constructed in this way in 
\cite{FF} (although the correct form of the equation itself did not appear until \cite{CH}). 
In the case at hand, with the vector 
$\bu=\left(\du, \dv \right)^T$, 
we apply this construction to the compatible Hamiltonian pairs listed in Theorem 
\ref{thH1}. Since the pairs of operators $\cJ^{(i)},\cH^{(i)}$ for $i=1,\ldots,6$ in the six cases  above depend linearly  on arbitrary constant parameters, in each case 
we have a lot of  freedom 
to obtain different
compatible pairs, by fixing the  constants in the operator $\cJ^{(i)}$ to get  $\cJ$, and taking linear combinations of 
$\cJ^{(i)}$ and $\cH^{(i)}$ with different constants to get $\cH$.

From case (i), we get the  integrable equation
\begin{eqnarray*}
 \left(\begin{array}{c} \du_t\\ 
\dv_t\end{array}\right)=\left(\cH^{(1)}+\left(\begin{array}{cc} c_1 \dd-c_2 
\dd^3& c_3\dd\\ c_3\dd & \!\!\!\! c_5 \dd\end{array} \right) \right) 
\left(\begin{array}{cc} 0
& \!\!\!\!\!\dd - \dd^2 \\ \dd + \dd^2 & 0
\end{array} \right)^{-1}\left(\begin{array}{c} \du_x\\ 
\dv_x\end{array}\right).
\end{eqnarray*}
Letting
\begin{eqnarray}\label{pquv}
 \du=(1-\dd)\pu, \quad \dv=(1+\dd)\pv,
\end{eqnarray}
 it follows that 
\begin{eqnarray}\label{eqh1}
 \left\{\begin{array}{l}(1-\dd)\pu_t=a_1\dd\pv 
(2 -\dd)\pu+c_3 \pu_x+c_1\pv_x-c_2\pv_{xxx}\\(1+\dd)\pv_t=\frac{a_1}{2}\dd 
(2 +\dd)\pv^2+c_3 \pv_x+c_5 \pu_x\end{array}\right. \qquad (a_1 c_2\neq 0).
\end{eqnarray}
In the same way, from cases (ii) and (iii) we get two pairs of integrable equations, given by 
\begin{eqnarray}
&& \left\{\begin{array}{l}(1-\dd)\pu_t=\frac{a_8}{2}\dd 
(2 -\dd)\pu^2+c_3 \pu_x+c_1\pv_x\\(1+\dd)\pv_t=a_8\dd\pu 
(2 +\dd)\pv+c_3 \pv_x+c_5 \pu_x-c_6 \pu_{xxx}\end{array}\right. 
\qquad (a_8 c_6\neq 0),\label{eqh2}\\
&& \left\{\begin{array}{l}(1-\dd)\pu_t=a_1\dd\pv 
(2-\dd)\pu+\frac{a_8}{2}\dd(2-\dd)\pu^2+c_3 \pu_x+c_1\pv_x\\
(1+\dd)\pv_t=\frac{a_1}{2}\dd(2+\dd)\pv^2+a_8\dd\pu 
(2 +\dd)\pv+c_3 \pv_x+c_5 \pu_x\end{array}\right. \ (a_1a_8 
\neq 0),\label{eqh3}
\end{eqnarray}
respectively. 
Notice that system (\ref{eqh2}) is the same as (\ref{eqh1}), upon swapping 
dependent variables $\pu\leftrightarrow\pv$ and sending $x\to -x$; they do not belong in the list in the previous section 
 since they include third derivatives, 
putting them  outside the family (\ref{chgensys}). In fact, the system (\ref{eqh1}) can be seen to be 
a reduction of  Example 2 on p.97 of 
\cite{strachan} by setting the parameters $h=0$, $f=1$ and performing a Galilean transformation.  It is also worth pointing out that it is possible to relax our elimination conditions slightly and still obtain interesting bi-Hamiltonian systems; for instance, setting $c_3=c_5=c_6=0$ in (\ref{eqh2}) or $a_1=c_3=c_5=0$ in  (\ref{eqh3}), with $a_8=c_1=1$ in both cases, gives Falqui's system (\ref{falqui}), which is  almost triangular (it would be with $c_1=0$).  

For the system (\ref{eqh3}), if we take $c_1=c_5=0$ and $c_3=1$ and rescale $\pu$ 
and $\pv$, we get the system (\ref{sysq2}) in Theorem \ref{th2}. Thus we arrive 
at the following result:
\begin{Cor}\label{cor1}
Define $\du$ and $\dv$ as in (\ref{pquv}). System (\ref{sysq2}) is 
bi-Hamiltonian, having the form 
 \begin{eqnarray*}
 \bu_t=\cH_1 \delta \rho_2=\cH_2 \delta\rho_1,
 \end{eqnarray*}
where the compatible Hamiltonian operators $\cH_1$ and $\cH_2$ are given by
 \begin{eqnarray*}
 \cH_1\!\!=\!\!\left(\!\!\begin{array}{cc}  0 \!&\!  \dd -\dd^2 \\ \dd +\dd^2 
 \!&\!  0\end{array} \!\!\right), \quad
   \cH_2\!\!=\!\!\left(\!\!\begin{array}{cc} 2\du 
\dd+2\dd \du \!&\! 2\dv\dd +  2\dd \du+\dd\\  2\dd \dv+ 2\du \dd+\dd \!&\!  
2\dv\dd+2\dd \dv 
\end{array} \!\!\right)
 \end{eqnarray*}
and the corresponding Hamiltonian functions are given by the densities $\rho_1=\pu \dv$ and 
$\rho_2=\pu \pv (2\du+2\dv+1)$.
\end{Cor}
\begin{Rem} To fix the notation, note that for the Hamiltonian $H=H[\bu]=\int \rho\, \rd x$ defined by the density $\rho$, we write 
the variational derivative as 
$$ 
\delta\rho = \frac{\delta H}{\delta \bu}=\left(\frac{\delta H}{\delta m}, \frac{\delta H}{\delta n}\right)^T, 
$$ 
 in the two-component case at hand. 
Also, recall that when $\rho$ (and hence $H$) is specified in terms of Miura-related variables ${\bf u}$, with the Miura map 
$\bu ={\cal M}({\bf u})$, 
the chain rule is  
$$ 
 \frac{\delta H}{\delta {\bf u}}={\cal M}_*({\bf u})^\dagger \frac{\delta H}{\delta \bu}, 
$$
where the star denotes the Fr\a'{e}chet derivative, and the dagger denotes the adjoint operator. 
\end{Rem}
\begin{Rem} Upon taking the linear combinations $\hat{u}=\frac{1}{2}(u-v)$, 
$\hat{v}=\frac{1}{2i}(u+v)$ and applying a Galilean transformation together with  
suitable rescalings, the system   (\ref{sysq2}) can be seen to be equivalent 
to Example 1 on p.97 of \cite{strachan} with parameters $h=\beta=0$. 
\end{Rem} 

For case (iv), if we let 
\begin{eqnarray}\label{pquv1}
 \du=(1-\dd^2)\pu; \quad \dv=(1-\dd^2)\pv ,
\end{eqnarray}
then we obtain the integrable equation 
\begin{eqnarray*}
 \left(\begin{array}{c} \du_t\\ 
\dv_t\end{array}\right)=\left(\cH^{(4)}+\left(\begin{array}{cc} c_1 \dd  
& 0 \\  0 & \!\!\!\! c_1 c_6 \dd\end{array} \right) \right) 
\left(\begin{array}{cc} \dd - \dd^3
& \!\!\!\!\! 0 \\  0  & c_6(\dd - \dd^3)
\end{array} \right)^{-1}\left(\begin{array}{c} \du_x\\ 
\dv_x\end{array}\right).
\end{eqnarray*}
in the explicit form
\begin{eqnarray}
&&  \left\{\begin{array}{l}\du_t=a_1 (2 \du \pu_x+\du_x \pu)+a_2(2 \dv 
\pu_x+\dv_x \pu+2 \du \pv_x+\du_x \pv)+\frac{a_6}{c_6} (2 \dv \pv_x+\dv_x 
\pv)+c_1 \pu_x\\
\dv_t=a_2 c_6 (2 \du \pu_x+\du_x \pu)\!+\!a_6 (2 \dv \pu_x+\dv_x
\pu+2 \du \pv_x+\du_x \pv)\!+\!\frac{a_8}{c_6}(2 \dv \pv_x+\dv_x \pv)+c_1 \pv_x 
\\ a_1 
a_6-a_2^2 c_6+a_2 a_8-a_6^2=0\ \mbox{and} \ 
c_6\neq 0\end{array}\right.\label{eqh6}
\end{eqnarray}
In particular, if we take either $a_6=0$ or $a_2=0$ then we get 
\begin{eqnarray}
 &&  \left\{\begin{array}{l}\du_t=a_1 (2 \du \pu_x+\du_x \pu)+a_2(2 \dv 
\pu_x+\dv_x \pu+2 \du \pv_x+\du_x \pv)\\
\dv_t=a_2 c_6 (2 \du \pu_x+\du_x \pu)+a_2 (2 \dv \pv_x+\dv_x
\pv) \end{array}\right. 
\qquad (a_2 c_6\neq 0),\label{eqh4}\\
&&  \left\{\begin{array}{l}\du_t=a_1 (2 \du \pu_x+\du_x 
\pu)+\frac{a_1}{c_6}(2 \dv\pv_x+\dv_x \pv)\\
\dv_t=a_1 (2 \dv \pu_x+\dv_x \pu+2 \du \pv_x+\du_x \pv)+\frac{a_8}{c_6} (2 \dv 
\pv_x+\dv_x
\pv) \end{array}\right. 
\qquad (a_1 c_6\neq 0),\label{eqh5} 
\end{eqnarray}
respectively. In fact, the latter  two systems 
are seen to be  the same
 by swapping 
 $\pu\leftrightarrow\pv$ and identifying 
parameters suitably. Taking $a_1=1$, $a_8=0$ and $c_6=-1$ in (\ref{eqh5}),
we get the two-component CH system (26) in \cite{Qu}.

For case (v), we let $\du=(1-\dd^2)\pu$. Then we get the 
integrable system 
\begin{eqnarray}
 &&  \left\{\begin{array}{l}\du_t=a_1 (2 \du \pu_x+\du_x \pu)+\frac{a_1}{c_5}\dv 
\dv_x+\frac{c_3}{c_5} 
\dv_x+c_1 \pu_x\\
\dv_t=c_3 \pu_x+a_1(\dv \pu_x+\dv_x\pu)  \end{array}\right. 
\qquad (a_1 c_5\neq 0).\label{eqh7}
\end{eqnarray}
For case (vi), we let $\dv=(1-\dd^2)\pv$. Then we get the  
integrable  system  
\begin{eqnarray}
 &&  \left\{\begin{array}{l}\du_t=c_3 \pv_x+a_8(\du 
\pv_x+\du_x\pv)\\
\dv_t=a_8 (2 \dv \pv_x+\dv_x \pv)+\frac{a_8}{c_1} \du \du_x+\frac{c_3}{c_1} 
\du_x +c_5\pv_x\end{array}\right. 
\qquad (a_8 c_1\neq 0).\label{eqh8}
\end{eqnarray}
Similarly to before, equations (\ref{eqh7}) and (\ref{eqh8}) are seen to be the same by 
swapping the dependent variables.
After taking $c_1=0$, $c_3=0$ and rescaling suitably, the system (\ref{eqh7}) becomes 
the known two-component CH equation (\ref{CLZ1}) from \cite{CLZ}.

With a change of notation,  the transformation (\ref{1miu}) presented in the introduction is 
\begin{eqnarray}\label{2t2}
 \pu=\bbu+\bbv, \quad \dv=(1-\dd)\bbu+(1+\dd)\bbv ,
\end{eqnarray}
which implies that
\begin{eqnarray}\label{2ti}
(1-\dd)\bbu=\frac{1}{2}\dd^{-1}(\du-(1-\dd)\dv), \quad (1+\dd)\bbv=\frac{1}{2}\dd^{-1}((1+\dd)\dv-\du) .
\end{eqnarray}
Thus equation (\ref{eqh7}) when $c_5=-1$, $c_3=1$ and $c_1=2$ becomes
\begin{eqnarray}
 &&  \left\{\begin{array}{l}(1-\dd)\bbu_t=a_1(2\bbu \bbu_x-\bbu \bbu_{xx}-\bbu_x^2+2 \bbu_x \bbv-\bbu_{xx}\bbv)+\bbu_x\\
(1+\dd)\bbv_t=a_1 (2\bbu \bbv_x+\bbu \bbv_{xx}+2 \bbv \bbv_x+\bbv \bbv_{xx}+\bbv_x^2)+\bbv_x\end{array}\right. 
\quad (a_1 \neq 0),\label{eqh77}
\end{eqnarray}
which is system (\ref{sysq1}) when $a_1=2$.
Thus we obtain the bi-Hamiltonian structure of system (\ref{sysq1}) by using the result for equation (\ref{eqh7}), as follows:
\begin{Cor}\label{cor12}
Define $\du$ and $\dv$ as in (\ref{pquv}). System (\ref{sysq1}) is a 
bi-Hamiltonian system, 
given by 
 \begin{eqnarray*}
\bu_t=\cH_1 
\delta \rho_2=\cH_2 \delta\rho_1,
 \end{eqnarray*}
where the compatible Hamiltonian operators $\cH_1$ and $\cH_2$ are given by
 \begin{eqnarray*}
&& \cH_1\!\!=\!\!-\frac{1}{2}\left(\!\!\begin{array}{cc}  0 \!&\!  \dd-1 \\ 
\dd+1
 \!&\!  0\end{array} \!\!\right), \\
 &&  \cH_2\!\!=\!\!\left(\!\!\begin{array}{cc} -\dd^{-1}\du_x-\du_x\dd^{-1} 
 \!&\! \du+\dv+\frac{1}{2}+\du_x \dd^{-1}-\dd^{-1}\dv_x\\  -(\du+\dv+\frac{1}{2})-\dv_x \dd^{-1}+\dd^{-1}\du_x \!&\!  
\dv_x\dd^{-1}+\dd^{-1} \dv_x 
\end{array} \!\!\right)
 \end{eqnarray*}
and the Hamiltonian functions are  given by the densities $\rho_1=2 \pu_x \dv$ and 
$\rho_2=2(\pv^2\du_x-\pu^2\dv_x-\pu^2\pv_x+\pv^2\pu_x+\pu_x\pv)$.
\end{Cor}
\begin{prf} To clarify the notation, we put hats on all variables in equation (\ref{eqh7}), that is, we write 
$\hat{\du}, \hat{\dv}$ etc. Thus the transformation (\ref{2ti}) becomes
$
 \du=\frac{1}{2}\dd^{-1}(\hat{\du}-(1-\dd)\hat{\dv})$, $\dv=\frac{1}{2}\dd^{-1}((1+\dd)\hat{\dv}-\hat{\du})$, 
so that 
\begin{eqnarray}\label{tra2}
\hat{\dv}=\du+\dv, \quad \hat{\du}=(1+\dd)\du+(1-\dd)\dv.
\end{eqnarray}
With $a_1=2$, $c_5=-1$, $c_3=1$ and $c_1=2$ in equation (\ref{eqh7}), 
the compatible Hamiltonian operators 
are
\begin{eqnarray*}
 \cH\!=\!\left(\!\!\!\begin{array}{cc} 2 (\hat{\du} \dd+\dd \hat{\du})+2 \dd & 
2 \hat{\dv} \dd +\dd \\
 2 \dd \hat{\dv} +\dd &0
\end{array} \!\!\!\right),  \quad \cJ\!=\!\left(\!\!\!\begin{array}{cc}  
\dd-\dd^3 & 0 \\0& - \dd
\end{array} \!\!\!\right).
\end{eqnarray*}
Under the transformation (\ref{tra2}), the Hamiltonian operator $\cJ$ is sent to 
\begin{eqnarray*}
\left(\!\!\!\begin{array}{cc}1+\dd&1-\dd\\1&1\end{array}\!\!\!\right)^{-1}
\!\!\!\cJ 
\left(\!\!\!\begin{array}{cc}1-\dd&1\\1+\dd&1\end{array}\!\!\!\right)^{-1}\!\!\!
=\frac{1}{4}\left(\!\!\!\begin{array}{cc}\dd^{-1}&1-\dd^{-1}\\-\dd^{-1}
& 1+\dd^{-1}\end{array} \!\!\!\right)
\!\!\cJ 
\left(\!\!\!\begin{array}{cc}-\dd^{-1}&\dd^{-1}\\1+\dd^{-1}&1-\dd^{-1}
            \end{array}
 \!\!\!\right) \!\!=\cH_1
\end{eqnarray*}
and $\cH$ is transformed to $\cH_2$ in the statement.
\end{prf}
\begin{Rem}
The inverse operator of Hamiltonian operator $2\cH_2$ in Corollary 
\ref{cor12} is of the form
 \begin{eqnarray*}
\!\!\left(\!\!\begin{array}{cc} 
\frac{2\dv_x}{(1+2\du+2\dv)^2}\dd^{-1}+\dd^{-1} 
\frac{2\dv_x}{(1+2\du+2\dv)^2}
 \!&\! -\frac{1}{1+2\du+2\dv}+\frac{2\dv_x}{(1+2\du+2\dv)^2}\dd^{-1}-\dd^{-1} 
\frac{2\du_x}{(1+2\du+2\dv)^2}\\  
\frac{1}{1+2\du+2\dv}-\frac{2\du_x}{(1+2\du+2\dv)^2}\dd^{-1}+\dd^{-1} 
\frac{2\dv_x}{(1+2\du+2\dv)^2}\!&\!  
-\frac{2\du_x}{(1+2\du+2\dv)^2}\dd^{-1}-\dd^{-1} 
\frac{2\du_x}{(1+2\du+2\dv)^2}
\end{array} \!\!\right)
 \end{eqnarray*}
Thus the local symmetries for system (\ref{sysq1}) can be generated
by the recursion operator $\cH_1 \cH_2^{-1}$.
\end{Rem}

\subsection{Compatible quadratic Hamiltonian operators}\label{sec42}
In this section, we consider  antisymmetric differential operators that are quadratic
in the dependent variables $\du$ and $\dv$, instead of linear as  in the 
previous subsection.
We assume that they are of the form
\begin{eqnarray}\label{HHc}
 &&\cH=\left(\begin{array}{cc} \cH_{11} & 
\cH_{12} \\
-\cH_{12}^\dagger &\cH_{22}
\end{array} \right)
+b_{14} \left( \begin{array}{c}\du_x\\ 
\dv_x\end{array}\right) \dd^{-1} \left( \begin{array}{cc}\du_x &
\dv_x\end{array}\right),
\end{eqnarray}
where, as before, $\dagger$ denotes the adjoint operator, with 
\begin{eqnarray*}
&&\cH_{11}= b_1 \du \dd \du+b_2 (\du \dd \dv +\dv \dd 
\du)+ b_3 \dv \dd \dv;\\
&&\cH_{12}=(b_4 \du^2+b_5 \du \dv +b_6 \dv^2)\dd+b_7 \du \du_x+b_8 \du 
\dv_x+b_9\du_x \dv+b_{10} \dv \dv_x;\\
&&\cH_{22}=b_{11} \du \dd \du+b_{12} (\du \dd \dv +\dv \dd 
\du) +b_{13}\dv \dd \dv 
\end{eqnarray*}
and $b_i, i=1,\ldots, 14$ being constants.

\begin{The}\label{thH2}
Let the operators $\cH$ and $\cJ$ be given by 
(\ref{HHc}) and  
(\ref{conH}), respectively.
\begin{itemize}
 \item Assume that $c_4= 1$. There is only one non-trivial CH Hamiltonian pair:
 \begin{eqnarray}
&& \cH^{(c)}\!\!=\!\!b_1\left(\!\!\!\begin{array}{cc}  \du \dd \du & 
 \dv \dd \du \\
\du \dd \dv  &  \dv \dd \dv \!\!\!
\end{array} \right)\!\!-\!\! b_1\left(\!\!\! \begin{array}{c}\du_x\\ 
\dv_x\end{array}\!\!\!\right)\!\! \dd^{-1}\!\! \left(\!\!\! 
\begin{array}{cc}\du_x &
\dv_x\!\!\!\end{array}\right)\!=\!b_1 \dd \left(\!\!\! \begin{array}{c}\du\\ 
\dv\end{array}\!\!\!\right)\!\! \dd^{-1}\!\! \left(\!\!\! \begin{array}{cc}\du &
\dv\!\!\!\end{array}\right)\dd,\label{cubich} \\
&&\cJ^{(c)}=\left(\!\!\!\begin{array}{cc} 
c_1 \dd
&\!\!\! c_3 \dd - \dd^2 \\
 c_3 \dd + \dd^2 & c_5 \dd 
\end{array}\!\!\! \right),  b_1 \neq 0; \label{cubicj}
 \end{eqnarray}
\item Assume that $c_4= 0$, $c_2=1$ or $c_4=c_2=0$, $c_6=1$, and the parameter 
$c_3$ is arbitrary. 
There are no non-trivial CH 
Hamiltonian pairs.
\end{itemize}
\end{The}
\begin{prf}
We prove this statement in the same way as we did for Theorem 
\ref{thH1}. Due to the large degree of similarity, we avoid tedious 
repetition and only write down the necessary steps and results.
The    operator $\cH$ is compatible with 
the  Hamiltonian operator $\cJ$ if and only if 
 the constants in $\cH$ and $\cJ$ 
satisfy an overdetermined algebraic system of the same type as (\ref{sys2}).
When $c_4=1$, we solve it and obtain only one solution after applying our 
elimination requirements: the nonzero constants in 
(\ref{HHc}) should satisfy 
$ -b_{14}=b_9=b_5=b_{13}=b_1$, and $c_2=c_6=0$. We denote the operators 
$\cH$ and $\cJ$ under the 
above constraints by $\cH^{(c)}$ and $\cJ^{(c)}$. By direct computation, 
 we are able to show the operator $\cH^{(c)}$ is Hamiltonian, and thus we 
obtain the  Hamiltonian pair in the statement. For other cases, there are no 
solutions for the above system after applying our elimination 
requirements.
\end{prf}
For the Hamiltonian pair given by (\ref{cubich}) and (\ref{cubicj}) we can immediately write 
down the integrable two-component equation
\begin{eqnarray*}
 \left(\begin{array}{c} \du_t\\ 
\dv_t\end{array}\right)=\left(\cH^{(c)}+\left(\begin{array}{cc} c_1 \dd& c_3 
\dd\\c_3 \dd & \!\!\!\! c_5 \dd\end{array} \right) \right) 
\left(\begin{array}{cc} 0
& \!\!\!\!\!\dd \!-\! \dd^2 \\ \dd \!+\! \dd^2 & 0
\end{array} \right)^{-1}\left(\begin{array}{c} \du_x\\ 
\dv_x\end{array}\right).
\end{eqnarray*}
We introduce the same notation for  $\pu$ and $\pv$ as in (\ref{pquv}). It 
follows that 
\begin{eqnarray}\label{eqh9}
 \left\{\begin{array}{l}(1-\dd)\pu_t=\frac{b_1}{2}\dd\pv 
(2 -\dd)\pu^2+c_3 \pu_x+c_1\pv_x\\(1+\dd)\pv_t=\frac{b_1 }{2}\dd \pu
(2 +\dd)\pv^2+c_5 \pu_x+c_3 \pv_x\end{array}\right. \qquad (b_1 \neq 0).
\end{eqnarray}
For equation (\ref{eqh9}), if we take $c_1=c_5=0$ and $c_3=1$ and rescale $\pu$ 
and $\pv$, then  we get the system (\ref{sysc2}) in Theorem \ref{th3}. Thus we have 
the following result.
\begin{Cor}\label{cor2}
Define $\du$ and $\dv$ as in (\ref{pquv}). System (\ref{sysc2}) is a 
bi-Hamiltonian system, that is, it takes the form 
 \begin{eqnarray*}
 \bu_t=\cH_1 
\delta \rho_2=\cH_2 \delta\rho_1,
 \end{eqnarray*}
where the  compatible Hamiltonian operators $\cH_1$ and $\cH_2$ are given by
 \begin{eqnarray*}
\cH_1\!\!=\!\!\left(\!\!\begin{array}{cc}  0 \!&\!  \dd -\dd^2 \\ \dd +\dd^2 
 \!&\!  0\end{array} \!\!\right), \quad
\cH_2\!\!=\!\!\left(\!\!\!\begin{array}{cc}  2\du \dd \du & 
2 \dv \dd \du+\dd \\
2 \du \dd \dv+\dd  &  2 \dv \dd \dv \!\!\!
\end{array} \right)\!\!-\!\! 2 \left(\!\!\! \begin{array}{c}\du_x\\ 
\dv_x\end{array}\!\!\!\right)\!\! \dd^{-1}\!\! \left(\!\!\! 
\begin{array}{cc}\du_x &
\dv_x\!\!\!\end{array}\right)
 \end{eqnarray*}
and the corresponding Hamiltonian functions are specified by the densities $\rho_1=\pu \dv$ and 
$\rho_2=\pu^2 \pv \dv+\pu \pv$.
\end{Cor}
Notice that we did not get system (\ref{sysc1}) in Theorem \ref{th3}. This is 
due to the assumptions we made on the Hamiltonian operators. Indeed, it is also 
bi-Hamiltonian, but does not have a Hamiltonian operator of the form (\ref{HHc});
this shows that the classification using Hamiltonian pairs is not equivalent to 
the symmetry approach. Here we just state the relevant result without proof, since the 
proof uses the same method as for Theorem \ref{thH1}.
\begin{Pro}\label{pro}
Define $\du$ and $\dv$ as in (\ref{pquv}). System (\ref{sysc1}) is a 
bi-Hamiltonian system, that is,
 \begin{eqnarray*}
\bu_t=\cH_1 
\delta \rho_2=\cH_2 \delta\rho_1,
 \end{eqnarray*}
where the compatible Hamiltonian operators $\cH_1$ and $\cH_2$ are given by
 \begin{eqnarray*}
&& \cH_1\!\!=\!\!\left(\!\!\begin{array}{cc}  0 \!&\!  \dd -1 \\ \dd +1 
 \!&\!  0\end{array} \!\!\right), \\ 
&&
\cH_2\!\!=\!\!\left(\!\!\!\begin{array}{cc}  \du \dd^{-1}\du_x+\du_x 
\dd^{-1}\du & 
-\du \dv-\frac{1}{2} +\du \dd^{-1}\dv_x -\du_x\dd^{-1}\dv\\
\du\dv+\frac{1}{2}-\dv \dd^{-1}\du_x+\dv_x\dd^{-1}\du  &  
-\dv\dd^{-1}\dv_x-\dv_x\dd^{-1}\dv \!\!\!
\end{array} \right)
 \end{eqnarray*}
and the corresponding Hamiltonian densities are $\rho_1=2\pu \dv_x$ and 
$\rho_2=\pu^2 \pv \dv_x+\pu^2\pv_x\dv+\pu \pv_x$.
\end{Pro}
\begin{Rem}
The inverse of the  Hamiltonian operator $2\cH_2$ in Proposition
\ref{pro} takes the form
 \begin{eqnarray*}
\!\!-\left(\!\!\begin{array}{cc} 
\frac{2\dv_x}{(1+2\du\dv)^2}\dd^{-1} \dv+\dv\dd^{-1} 
\frac{2\dv_x}{(1+2\du\dv)^2}
 \!&\! -\frac{1}{1+2\du\dv}+\frac{2\dv_x}{(1+2\du\dv)^2}\dd^{-1}\du-\dv\dd^{-1} 
\frac{2\du_x}{(1+2\du\dv)^2}\\  
\frac{1}{1+2\du\dv}-\frac{2\du_x}{(1+2\du\dv)^2}\dd^{-1}\dv+\du \dd^{-1} 
\frac{2\dv_x}{(1+2\du\dv)^2}\!&\!  
-\frac{2\du_x}{(1+2\du\dv)^2}\dd^{-1} \du-\du \dd^{-1} 
\frac{2\du_x}{(1+2\du\dv)^2}
\end{array} \!\!\right)
 \end{eqnarray*}
Thus the local symmetries for system (\ref{sysc1}) can be generated
using  the recursion operator $\cH_1 \cH_2^{-1}$.
\end{Rem}

It follows from Corollary \ref{cor1} and Corollary \ref{cor2} that both systems 
possess the same Hamiltonian operator $\cH_1$ (in fact, $\cJ^{(3)}=\cJ^{(c)}$).
So both Hamiltonian operators $\cH^{(3)}$ 
and $\cH^{(c)}$ form Hamiltonian pairs with the same operator. We are 
able to directly verify that any linear combination of $\cH^{(3)}$ 
and $\cH^{(c)}$ is also Hamiltonian, and forms a Hamiltonian pair with 
$\cJ^{(3)}$.  Thus we can construct the integrable system 
\begin{eqnarray}\label{qucu}
\left\{\begin{array}{l}(1-\dd)\pu_t=a_1\dd\pv 
(2-\dd)\pu+\frac{a_8}{2}\dd(2-\dd)\pu^2+\frac{b_1}{2}\dd\pv 
(2 -\dd)\pu^2+c_3 \pu_x+c_1\pv_x , 
\\(1+\dd)\pv_t=\frac{a_1}{2}\dd(2+\dd)\pv^2+a_8\dd\pu 
(2 +\dd)\pv+\frac{b_1 }{2}\dd \pu
(2 +\dd)\pv^2+c_5 \pu_x+c_3 \pv_x ,\end{array}\right. 
\end{eqnarray}
which contains both equations (\ref{eqh3}) and (\ref{eqh9}).
If we take $c_1=c_5=0$, $c_3=1$, $a_8=2 \alpha, a_1=2 \beta,$ and $b_1=2\gamma$, 
then we get the system (\ref{sysm1}) in Theorem \ref{th4}. Thus we have 
the following result.
\begin{Cor}\label{cor3}
Define $\du$ and $\dv$ as in (\ref{pquv}). System (\ref{sysm1}) is 
bi-Hamiltonian, being given by 
 \begin{eqnarray*}
 \bu_t=\cH_1 
\delta \rho_2=\cH_2 \delta\rho_1,
 \end{eqnarray*}
where the compatible Hamiltonian operators $\cH_2$ and $\cH_1$ are given by
 \begin{eqnarray*}
 && \cH_2\!\!=\!\!\left(\!\!\begin{array}{cc} 2\beta(\du 
\dd+\dd \du)+2\gamma\du \dd \du \!&\! 2 \beta \dv\dd + 2 \alpha\dd \du+2\gamma 
\dv \dd \du+\dd\\ 2 \beta\dd \dv+ 2 \alpha\du \dd+2\gamma \du \dd \dv+\dd 
\!&\!  
2 \alpha(\dv\dd+\dd \dv )+2\gamma \dv \dd \dv
\end{array} \!\!\right)\\
&&\qquad - 2\gamma \left(\!\!\! \begin{array}{c}\du_x\\ 
\dv_x\end{array}\!\!\!\right)\!\! \dd^{-1}\!\! \left(\!\!\! 
\begin{array}{cc}\du_x &
\dv_x\!\!\!\end{array}\right),
\qquad \cH_1\!\!=\!\!\left(\!\!\begin{array}{cc}  0 \!&\!  \dd -\dd^2 \\ \dd 
+\dd^2 \!&\!  0\end{array} \!\!\right)
 \end{eqnarray*}
and the corresponding Hamiltonian densities are $\rho_1=\pu \dv$ and 
$\rho_2=\pu \pv(2 \alpha \du+2\beta \dv+\gamma \pu\dv+1)$.
\end{Cor}

The same situation arises for systems (\ref{sysq1}) and (\ref{sysc1}), upon 
comparing Corollary \ref{cor12} to Proposition \ref{pro}.  We present 
 the result immediately, as follows.
\begin{Cor}
 Define $\du$ and $\dv$ as in (\ref{pquv}). System (\ref{sysm2}) takes the 
bi-Hamiltonian form 
 \begin{eqnarray*}
 \bu_t=\cH_1 
\delta \rho_2=\cH_2 \delta\rho_1,
 \end{eqnarray*}
where the compatible Hamiltonian operators $\cH_1$ and $\cH_2$ are given by
 \begin{eqnarray*}
&&\cH_1\!\!=\!\!\left(\!\!\begin{array}{cc}  0 \!&\!  \dd -1 \\ \dd +1 
 \!&\!  0\end{array} \!\!\right), \\
&&\cH_2\!\!=\!\!\gamma\left(\!\!\!\begin{array}{cc}  \du \dd^{-1}\du_x+\du_x 
\dd^{-1}\du & 
-\du \dv+\du \dd^{-1}\dv_x -\du_x\dd^{-1}\dv\\
\du\dv-\dv \dd^{-1}\du_x+\dv_x\dd^{-1}\du  &  
-\dv\dd^{-1}\dv_x-\dv\dd^{-1}\dv_x\!\!\!
\end{array} \right)\\
&&\quad +\!\!\left(\!\!\begin{array}{cc} \beta(\dd^{-1}\du_x+\du_x\dd^{-1}) 
 \!&\! -\alpha(\du+\du_x \dd^{-1})-\beta(\dv-\dd^{-1}\dv_x)-\frac{1}{2}\\  
\alpha(\du-\dd^{-1}\du_x)+\beta(\dv+\dv_x \dd^{-1})+\frac{1}{2} \!&\!  
-\alpha(\dv_x\dd^{-1}+\dd^{-1} \dv_x )
\end{array} \!\!\right)
 \end{eqnarray*}
and the corresponding Hamiltonian densities are $\rho_1=2\pu \dv_x$ and 
$$\rho_2=\gamma \pu^2 (\pv \dv_x+\pv_x\dv)-\beta \pv^2 (\du_x+\pu_x)+\alpha 
\pu^2(\dv_x+\pv_x)+\pu \pv_x .$$
\end{Cor}

\section{Reciprocal links, 
Lax pairs and exact solutions} 

In this section we describe reciprocal transformations relating the coupled Camassa-Holm type systems 
to negative flows in other known integrable hierarchies. We also present Lax pairs, and provide 
some exact solutions in certain cases. 

Before we proceed, it is worth commenting on the linear terms appearing in the systems under consideration. 
It is necessary to include linear dispersion terms  in order to be able to apply the perturbative symmetry approach. 
However, given a system in the form (\ref{chgensysred}), we can rescale the dependent variables and time so that 
$u\to \epsilon^{-1}u$, 
$v\to \epsilon^{-1}v$, 
$t\to \epsilon^{d}t$, 
where $d$ is the common degree of $f$ and $g$ in $u,v$ and their derivatives,
and then take the limit 
$ \epsilon\to 0$, to obtain the system in the form 
$$ 
m_t =\hat{f}, \qquad 
n_t=\hat{g}, 
$$ 
where $m=u-u_x$, $n=v+v_x$ and $\hat{f},\hat{g}$ are homogeneous of degree $d$.  
In general, the latter system is not  isomorphic to the original system  (\ref{chgensysred}), although 
 this is the case for the first quadratic system (\ref{sysq1}). Indeed, if we perform 
 a combination of shifting the dependent variables with a Galilean transformation, that is 
\beq\label{shift} 
u\to u+u_0, \qquad v\to v+v_0, \qquad x \to x-ct, \qquad t\to t, 
\eeq 
then with $u_0=v_0$ and a suitable choice of $c$ it is possible to remove the $u_x$ and $v_x$ terms on the 
right-hand side of (\ref{sysq1}). However, for the second quadratic system (\ref{sysq2}) this is not the case, because applying 
(\ref{shift}) creates a mixture of $u_x$ and $v_x$ terms on the right-hand side of the system; and for the systems 
with cubic nonlinearity, applying (\ref{shift}) produces additional quadratic and linear terms.

\subsection{First quadratic system} 

We have already seen that the first quadratic system is related by a Miura map to the system (\ref{CLZ1}) 
of Chen-Liu-Zhang. This means that we can immediately obtain a Lax pair for  (\ref{sysq1}), 
by  using the results in  \cite{CLZ}. %
\begin{Pro}\label{quad1lax} 
The  first quadratic system  (\ref{sysq1}) has the  Lax representation 
$$
\begin{array}{rcl} 
\psi_{xx} & =& \left[\lambda^2\Big(2(m+n)+1\Big)^2-\lambda\Big(2(m+n)+1+2m_x-2n_x\Big)+\frac{1}{4}\right]\psi,\\
\psi_t&=&\,\left(\frac{1}{2\lambda}+2(u+v)+1\right)\psi_x-(u_x+v_x)\psi,
\end{array} 
$$
where  $m=u-u_x$, $n=v+v_x$.
\end{Pro}
In fact, as already mentioned, the linear dispersion terms can be removed from this particular system without 
taking any scaling limit, by using (\ref{shift}), 
and after rescaling time the system becomes (\ref{CLZ}), which can be written in the form 
$$ 
m_t=(um)_x+v(m+u)_x, \qquad n_t=(vn)_x+u(n+v)_x.
$$ 

In order to obtain solutions of the system, it is helpful to make use of the second equation of the system (\ref{CLZ1}), which is in 
conservation form, and leads to the introduction of new independent variables $X,T$ via the reciprocal transformation 
\beq\label{clzrt} 
\rd X = q\, \rd x +pq\, \rd t, \quad \rd T = \rd t, \qquad \text{with} \quad p=u+v, \quad q=m+n. 
\eeq 
As explained in  \cite{CLZ}, this change of independent variables transforms (\ref{CLZ1}) to the first negative flow of the AKNS 
hierarchy (which, at the level of the Lax pair, is equivalent to the classical Boussinesq hierarchy, up to a gauge transformation). 
Under the reciprocal transformation, we have the following system of four equations relating $u,v,m,n$: 
\beq \label{4sys} 
\begin{array}{rcl} 
m_T=(m+n)(m+v)u_X, &\qquad &(m+n)u_X=u-m, \\ 
n_T=(m+n)(n+u)v_X, & & (m+n)v_X=n-v.  
\end{array} 
\eeq 
Solutions of this system, as functions of $X,T$, lead to parametric solutions of the original system  (\ref{CLZ}). However, it turns out that it is more convenient to first obtain solutions of the second quadratic system, as described in the next subsection, and then exploit a Miura map between the two systems, rather than attempting to solve (\ref{4sys}) directly. 

\subsection{Second quadratic system} 

In this subsection we consider the second  quadratic system  (\ref{sysq2}) without linear dispersion terms, which (after rescaling $t$) can be written as 
\beq\label{quasys2} 
M_t=\Big( (U+V)M +UV\Big)_x, \qquad N_t=\Big( (U+V)N +UV\Big)_x,
\eeq 
with $M=U-U_x$, $N=V+V_x$, where all the dependent variables are given upper case letters to distinguish them from the 
variables in the first   quadratic system. 
The need to make this distinction here  is due to the following result.

\begin{Pro}\label{qmiura} 
A solution of the first  quadratic system  (\ref{CLZ}) gives rise to a solution of the second quadratic system (\ref{quasys2}) via the Miura map 
\beq\label{qmiu} 
U=u+v+v_x, \qquad V=-v_x
\eeq 
\end{Pro} 
\begin{prf}
From (\ref{qmiu}) it follows that $U=u+n$ and $U+V=u+v$, so that  
$$ 
M=m+(1-D_x)n, \qquad N=-n_x,
$$ 
which gives 
$$
M-N=m+n, \qquad M+N=m+n-2n_x.
$$
Upon taking the time derivative of the latter two equations and using (\ref{CLZ}), 
we see that the difference and sum of $M$  and $N$ evolve according to 
\beq\label{MNsys} 
(M-N)_t=\Big((U+V)(M-N)\Big)_x, \qquad (M+N)_t=\Big((U+V)(M+N)+2UV\Big)_x,
\eeq 
which is equivalent to (\ref{quasys2}). 
\end{prf}

From the above, we see that the system  (\ref{quasys2}) is intermediate between (\ref{CLZ}) and (\ref{CLZ1}), and we 
can write the Miura map from  (\ref{quasys2}) to (\ref{CLZ1}) directly as 
\beq\label{miu2}
p=U+V, \qquad q=M-N.
\eeq
By taking the Lax pair in \cite{CLZ}, or by shifting/scaling the coefficients of the Lax pair in Proposition \ref{quad1lax} and using 
(\ref{qmiu}), we immediately have the following. 
\begin{Pro}\label{quad2lax}
The second quadratic system  (\ref{quasys2})  has the  Lax representation 
$$
\begin{array}{rcl} 
\psi_{xx} & =& \left[\lambda^2 (M-N)^2-\lambda (M+N+M_x-N_x)+\frac{1}{4}\right]\psi,\\
\psi_t&=&\,\left(\frac{1}{2\lambda}+U+V\right)\psi_x-\frac{1}{2}(U_x+V_x)\psi,
\end{array} 
$$
where $M=U-U_x$, $N=V+V_x$.
\end{Pro}
Next, observe that  the first equation in (\ref{MNsys}) is just the conservation law $q_t=(pq)_x$. This means that 
the same 
reciprocal transformation (\ref{clzrt}) can be used to link  (\ref{quasys2}) to the first negative AKNS flow. The 
equations (\ref{MNsys}) and the relations $M=U-U_x$, $N=V+V_x$ are transformed to 
\beq \label{4sys2} 
\begin{array}{rcl} 
(q^{-1})_T+p_X=0, &\qquad &qU_X=U-M, \\ 
\Big((M+N)q^{-1}\Big)_T=(2UV)_X, & & qV_X=N-V.  
\end{array} 
\eeq 
Upon adding and subtracting the equations that involve only $X$ derivatives, and using (\ref{miu2}), we see that the relations 
\beq\label{MNrels} 
(M+N)q^{-1}=pq^{-1}-(U-V)_X, \qquad U-V=q(1+p_X) 
\eeq 
hold. With the  introduction of a potential $f(X,T)$ into the conservation law for $q^{-1}$, 
such that $q^{-1}=f_X$, $p=-f_T$, it is possible to use   (\ref{miu2}) and (\ref{MNrels}) to express $U,V$ purely 
in terms of derivatives of $f$.  Moreover, all of the terms in the conservation law for $(M+N)q^{-1}$ in (\ref{4sys2}) can also be 
rewritten in terms of $f$, to yield a single equation for this potential, namely 
\beq\label{forig} 
\left( \Big(\frac{f_{XT}-1}{f_X}\Big)_X-f_Xf_T\right)_T=\frac{1}{2}\left(f_T^2-\frac{(f_{XT}-1)^2}{f_X^2}\right)_X.
\eeq 
The latter equation is equivalent to equation (2.16) in \cite{CLZ}; below we rewrite it  in a form which makes it more 
 easily identifiable as such. 
\begin{The} \label{2quadrt} 
Let $f(X,T)$ be a solution of the equation 
$$ 
f_X^2f_{XXTT}-f_Xf_{XX}f_{XTT}-f_Xf_{XT}f_{XXT}-f_X^4f_{TT}+f_{XX}f_{XT}^2
-2f_X^3f_{T}f_{XT}-f_{XX}=0.
$$
Then taking 
\beq\label{UVp} 
U=\frac{1}{2}\Big(-f_T-(f_{XT}-1)f_X^{-1}\Big), \qquad 
V=\frac{1}{2}\Big(-f_T+(f_{XT}-1)f_X^{-1} \Big) 
\eeq
and 
$$
x=f(X,t)
$$ 
gives a solution $(U(x,t),V(x,t))$ of  
 the system  (\ref{quasys2}) in parametric form. 
\end{The} 
\begin{Cor} A solution $(u(x,t),v(x,t))$ of  
 the system  (\ref{CLZ}) is given in parametric form by taking 
$$ 
 u=\frac{1}{2}\Big(-X-f_{T}  -\int f_Xf_T\, \rd X \Big), 
\qquad 
v=\frac{1}{2}\Big(X-f_{T}  +\int f_Xf_T\, \rd X \Big).
$$ 
\end{Cor}
{\bf Proof of Corollary:} Applying the reciprocal transformation (\ref{clzrt}) to the second equation 
in (\ref{qmiu}) yields   $v_X=-q^{-1}V=-f_XV$. The expression for $v$ then follows by using the formula for $V$ 
in (\ref{UVp}) and integrating with respect to $X$ (which leaves a function of time unspecified); $u$ is then 
found by noting that $u+v=U+V=p=-f_T$. \qed
\begin{figure}
\centering
\begin{subfigure}{.5\textwidth}
  \centering
  \includegraphics[width=\linewidth]{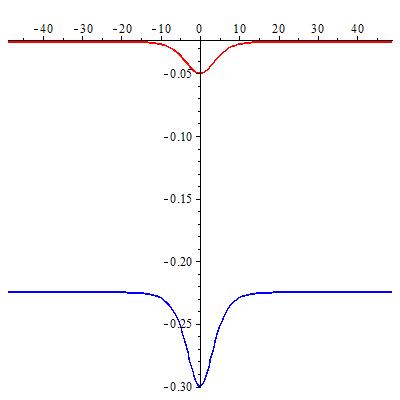}
  \caption{Dark soliton case: $k=1/2$, $\delta=1$, $r_0=4$ $\quad$ with $c=3/4$, $\rC=1/10$.}
\end{subfigure}%
\begin{subfigure}{.5\textwidth}
  \centering
  \includegraphics[width=\linewidth]{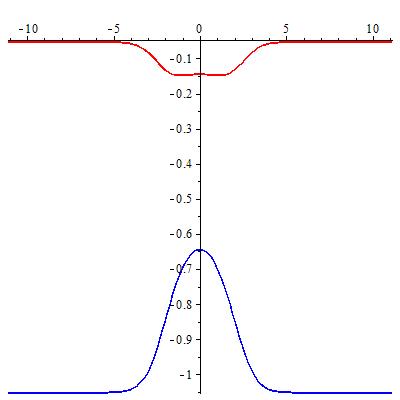}
  \caption{Bright soliton case: $k=1/2$, $\delta=1$, $r_0=2$ $\quad$ with $c=9\sqrt{10}/20$, $\rC=1/\sqrt{10}$.} 
\end{subfigure}
\caption{Parametric plots of $U$ (red) and $V$ (blue) showing dark/bright solitons.}
\label{UVdb}
\end{figure}

\noindent {\bf Example: travelling waves.} Travelling waves of the system 
(\ref{quasys2}) depend on $x,t$ via the combination $z=x-ct$, where $c$ is the wave velocity. 
They are obtained in parametric form by taking
$$
z=\hat{f}(Z), \qquad f(X,T) =\hat{f}(Z)+cT,\qquad Z=X-\rC T,  
$$
which gives solutions of (\ref{forig}) corresponding to travelling waves with velocity $\rC$ in the reciprocally transformed 
system (\ref{4sys2}). If we set $\rho =\hat{f}'$, then (\ref{forig})  becomes an ordinary differential 
equation of third order for $\rho$, and after integrating twice this yields 
\beq\label{rhoell} 
(\rho ')^2= \rho^4-2c\rC^{-1} \rho^3+K_2\rho^2+K_1\rho+\rC^{-2}, 
\eeq  
where $K_1,K_2$ are arbitrary constants. The general solution of the latter equation is an elliptic function 
$\rho(Z)$. In general, from (\ref{UVp}), $U$ and $V$ are then given in parametric form in terms of $\rho(Z)$ 
and $\rho'(Z)$ according to 
\beq\label{UVrho}
U=\frac{1}{2}\left(\rC\rho -c+\frac{\rC\rho '+1}{\rho} \right), \qquad 
V=\frac{1}{2}\left(\rC\rho -c-\frac{\rC\rho '+1}{\rho} \right).
\eeq

Here we consider single soliton solutions, which are obtained by choosing the quartic in 
(\ref{rhoell}) to have a double root. In that case, the solutions take the form 
\beq\label{rhos} 
\rho(Z) = r_0 \pm\left( \frac{2\delta k(1-k^2)\mathrm{sinh}^2(\delta Z)}{1+(1-k^2)\mathrm{sinh}^2(\delta Z)}\right), 
\qquad 0<k<1, \eeq 
where the values of $\rC$ and $c$ are fixed by the choice of parameters $k$ and $r_0,\delta>0$. To be more precise,
 substituting the solution (\ref{rhos}) into (\ref{rhoell}) determines $C,c,K_1,K_2$, and $r_0$ must be chosen to ensure 
that $\rd z/\rd Z=\rho(Z)>0$ everywhere, in order for the parametric solution for $U,V$ to be single-valued. Upon integrating 
(\ref{rhos}), the similarity variable $z=x-ct$ is obtained as 
$$ 
z =r_0 Z\pm 2k \log\left(\frac{1+\mathrm{tanh}(\delta Z/2)}{1+\mathrm{tanh}(\delta Z/2)}\right) 
+ \log\left(\frac{1\mp 2k \mathrm{tanh}(\delta Z/2)+\mathrm{tanh}^2(\delta Z/2)}{1\pm 2k \mathrm{tanh}(\delta Z/2)+\mathrm{tanh}^2(\delta Z/2)}\right) , 
$$ up to shifting by an arbitrary constant. The field $\rho$ has the shape of a dark soliton (a wave of depression) when the 
plus sign is chosen in (\ref{rhos}), while with a minus sign it is a bright soliton; in Figure \ref{UVdb} the corresponding 
fields $U,V$ given by (\ref{UVrho}) are plotted in these two different cases.

\subsection{First  cubic system} 

For simplicity, we consider the system (\ref{sysc1}) in the absence of linear terms on the right-hand sides, in which case 
(with suitable scaling) it  can be written as 
\beq\label{cusys1}  
m_t=v(um)_x, \qquad n_t = u(vn)_x, 
\qquad 
\text{with} \quad m=u-u_x, \, n= v+v_x .
\eeq
In that case, it is useful to consider the first non-trivial 
 symmetry of the system, which  (up to rescaling) takes the form 
\beq\label{cusymp1} 
u_\tau= \frac{m_x}{(mn)^2} , \qquad v_\tau= \frac{n_x}{(mn)^2} .
\eeq
The quantity $F=mn$ is a conserved density for both (\ref{cusys1}) and the latter symmetry, which satisfies 
\beq\label{Fttau} 
F_t=(uv\, F)_x, \qquad F_\tau = -G_x, \qquad \text{with} \quad F=mn, \, G=F^{-1}\Big(1+(\log(m/n))_x\Big). 
\eeq 
In order to find the Lax pair for the cubic system, it is helpful to consider a simultaneous reciprocal transformation 
in the independent variables $x,t,\tau$, by setting 
\beq\label{curttau} 
\rd X=F\, \rd x + uvF\,\rd t - G\rd \tau, \quad \rd T=\rd t, \quad \rd s =\rd \tau. 
\eeq 
 (Of course, this could be extended to include the whole hierarchy of symmetries of 
(\ref{cusys1}), but  the symmetry $\partial \tau$ is sufficient for our purposes.) 
The partial derivatives transform as $\partial_x=F\partial_X$, $\partial_t=\partial_T +uvF\partial_X$ and 
$\partial_\tau=\partial_s-G\partial_X$. 
To begin with, we identify the symmetry (\ref{cusymp1}) by introducing new dependent variables 
$$ 
p=\frac{1}{m}, \qquad q=\frac{1}{n}, 
$$ 
and find that under the 
reciprocal transformation (\ref{curttau}) it yields a system of derivative nonlinear 
Schr\"{o}dinger type, namely 
\beq \label{cll} 
p_s = -p_{XX} +2qpp_X, \qquad 
q_s = -q_{XX} +2pqq_X,
\eeq 
which is the Chen-Lee-Liu system \cite{cll}. For the latter system, 
we take the Lax pair in the form 
\beq\label{cllLax} \begin{array} {l} 
\Psi_X={\bf F} \Psi,\quad \Psi_s={\bf G}\Psi,  \qquad  
\text{with} \qquad   
{\bf F} =\left(\begin{array}{cc}\frac{1}{2}(\la+pq) & -q \\ 
p\la & - \frac{1}{2}(\la+pq) \end{array}\right), \\ 
{\bf G} =\left(\begin{array}{cc}\frac{1}{2}\la^2 +pq\la +\frac{1}{2}(pq_X-p_Xq+p^2q^2) & 
-q\la-q_X-pq^2 \\  p\la^2+(-p_X+p^2 q) \la & 
-\frac{1}{2}\la^2 -pq\la -\frac{1}{2}(pq_X-p_Xq+p^2q^2) \end{array}\right) .
\end{array} 
\eeq 
If the same reciprocal transformation  (\ref{curttau}) is applied to (\ref{cusys1}), then in terms of the variables $p,q,u,v$ 
we find a system given by two pairs of equations, that is 
\beq\label{rtcub1} 
\begin{array}{rclrcl} 
p_T & = & v-puv, \quad & q_T& =& -u+quv, \\ 
u_X & =& -q+upq, & v_X & = & p-vpq, \\  
\end{array} 
\eeq 
which is symmetrical under the involution 
\beq\label{invol} 
p\leftrightarrow -u, \qquad q\leftrightarrow -v, \qquad X\leftrightarrow -T. 
\eeq 
The latter system corresponds to a negative flow in the hierarchy of symmetries of the  Chen-Lee-Liu system \cite{cll}, 
and its Lax pair is found by taking the same $X$ part as in (\ref{cllLax}) and a $T$ part which is linear in the inverse 
of the spectral parameter $\la$. 
\begin{Pro} 
The system (\ref{rtcub1}) has the Lax pair 
\beq\label{rtcu1Lax} 
\Psi_X={\bf F} \Psi,\qquad \Psi_s={\bf H}\Psi, 
\eeq 
where ${\bf F}$ is as in (\ref{cllLax}), and 
$$ 
{\bf H}= \left(\begin{array}{cc}\frac{1}{2}(\la^{-1}+uv) & -u\la^{-1} \\ 
v & - \frac{1}{2}(\la^{-1}+uv) \end{array}\right). 
$$ 
\end{Pro} 
\begin{Rem} Upon taking the first component of the vector $\Psi$ to be $\psi_1=\sqrt{q}\phi$, the $X$ part of the Lax pair implies that the function $\phi$ is a solution of the energy-dependent Schr\"{o}dinger equation 
$$ 
\phi_{XX}=\Big(\frac{1}{4}\la^2 + U\la +V\Big)\phi, 
$$ 
where $U,V$ are certain functions of $p,q$ and their derivatives. This shows that 
the system (\ref{rtcub1}) is related by a Miura transformation to the first negative flow in the classical Boussinesq hierarchy.      
\end{Rem} 
\begin{Cor} The system  (\ref{cusys1})  has the Lax pair 
\beq\label{cu1Lax} 
\begin{array}{rcl} 
\left(\begin{array}{c}\psi_{1,x} \\ \psi_{2,x}\end{array}\right)&  =&\left(\begin{array}{cc} \frac{1}{2}( mn \la +1) & -m \\ 
n\la & -\frac{1}{2}( mn\la +1)  \end{array}\right) 
\left(\begin{array}{c}\psi_{1} \\ \psi_{2}\end{array}\right),  \\ 
\left(\begin{array}{c}\psi_{1,t} \\ \psi_{2,t}\end{array}\right)&  =&\left(\begin{array}{cc} \frac{1}{2}(uv mn\la +2uv+\la^{-1}) & -uvm -u\la^{-1}\\ 
uvn\la +v & -\frac{1}{2}(uv mn\la +2uv+\la^{-1})  \end{array}\right)  
\left(\begin{array}{c}\psi_{1} \\ \psi_{2}\end{array}\right). 
\end{array} 
\eeq 
\end{Cor} 
{\bf Proof of Corollary:} This follows immediately by applying the inverse of the  reciprocal transformation  (\ref{curttau}) to 
the vector wave function $\Psi = (\psi_1,\psi_2)^T$ in (\ref{rtcu1Lax}). \qed

As it stands, the system (\ref{rtcub1}) is not so easy to analyse from the point of view of obtaining solutions. However,  the 
dependent variables $u,v$ can be rewritten in terms of $p,q$ and their derivatives according to the expression 
\beq\label{uvex} 
\left(\begin{array}{c}u \\ v\end{array}\right)=\frac{1}{2w^2-w(\log r)_X} \left(\begin{array}{cc}  q_X-qw & -q \\ 
p_X+pw & -p  \end{array}\right) \left(\begin{array}{c}w_T\\ w_{XT}-2w\end{array}\right), 
\eeq   
where 
$$ 
w=pq, \qquad r=\frac{q}{p}. 
$$ 
Under the  reciprocal transformation  (\ref{curttau}), the conservation law for $F=w^{-1}$ 
becomes $w_T+(uv)_X=0$, and using (\ref{uvex}) the product $\Pi=uv$ can be rewritten purely in terms of $w$ and $r$, leading to a system for these two variables alone, namely 
\beq \label{rwsys} 
w_T +\Pi_X=0, \qquad (\log r)_T=2\Pi -\frac{\cal A}{\cal B}, 
\eeq
with 
$$ \Pi=\frac{w}{4}\left(\frac{{\cal A}^2}{{\cal B}^2}-\frac{w_T^2}{w^2}\right), 
\qquad {\cal A} = 2\frac{w_{XT}}{w}-\frac{w_Xw_T}{w^2}-4, \qquad 
{\cal B} =(\log r)_X-2w. 
$$  
The system (\ref{rwsys}) passes the Painlev\'{e} test with expansions around a movable singularity manifold 
$\varphi(X,T)=0$ having the two different leading order behaviours 
$w\sim \pm \varphi^{-1}$, $\log r\sim \mp \log\varphi$. Moreover, 
from a solution of this system one recovers $p,q$, and hence  
also $u,v$ from (\ref{uvex}), as functions of $X$ and $T$; via the reciprocal   transformation  (\ref{curttau}), this 
produces 
a solution of (\ref{cusys1}). 

\begin{The} \label{uvparam} 
Let $(f(X,T), r(X,T))$ be a solution of the system 
\beq\label{rfsys} 
\begin{array}{rcl} 
4f_T+f_X\Big(\hat{\cal A}^2\hat{\cal B}^{-2}-f_{XT}^2f_X^{-2}\Big) & = & 0,  \\ 
(\log r)_T + 2f_T +\hat{\cal A}\hat{\cal B}^{-1} & = & 0, 
\end{array} 
\eeq
with 
$$ 
\hat{\cal A} = 2\frac{f_{XXT}}{f_X}-\frac{f_{XX}f_{XT}}{f_X^2}-4, \qquad 
\hat{\cal B}=(\log r)_X-2f_X. 
$$  
Then taking $w (X,T)=f_X(X,T)$, $p=\sqrt{w/r}$, $q=\sqrt{wr}$  together with (\ref{uvex}), 
and setting $$x=f(X,t)$$ gives a  solution $(u(x,t),v(x,t))$
of the system (\ref{cusys1})
in parametric form. Equivalently, a parametric solution of (\ref{cusys1}) is obtained from a 
solution $(f(X,T), r^*(X,T))$  of the system 
\beq\label{rstarfsys} 
\begin{array}{rcl} 
4f_X+f_T\Big(({\cal A}^*)^2({\cal B}^*)^{-2}-f_{XT}^2f_T^{-2}\Big) & = & 0,  \\ 
(\log r^*)_X + 2f_X -{\cal A}^*({\cal B}^*)^{-1} & = & 0, 
\end{array} 
\eeq
with 
$$ 
{\cal A}^* = 2\frac{f_{XTT}}{f_T}-\frac{f_{XT}f_{TT}}{f_T^2}-4, \qquad 
{\cal B}^*=-(\log r^*)_T+2f_T, 
$$  
by taking $\Pi (X,T)=-f_T(X,T)$, $u=\sqrt{\Pi/r^*}$, $v=\sqrt{\Pi r^*}$. 
\end{The} 
\begin{prf} The quantity $f$ arises by introducing a potential in the first  equation in (\ref{rwsys}). 
The differential of the above formula for $x$ gives $\rd x = f_X (X,t)\, \rd X +f_T(X,t)\rd t 
=w\,\rd X -\Pi\, \rd T$, which follows by identifying $\Pi=-f_T$, ${\cal A}=\hat{\cal A}$ and ${\cal B}=\hat{\cal B}$, upon comparing the terms from the second equation in (\ref{rwsys}) with those in (\ref{rfsys}); this expression for $\rd x$  
corresponds precisely to  the inverse of the reciprocal transformation  (\ref{curttau}), as required. The same  parametric solution $(u(x,t),v(x,t))$, with 
$x=f(X,t)$, can be obtained in a different way by exploiting the symmetry (\ref{invol}). Indeed, 
from (\ref{rtcub1}), or by applying the involution to 
(\ref{uvex}),  the 
dependent variables $p,q$ can be rewritten in terms of $u,v$ and their derivatives as 
\beq\label{pqex} 
\left(\begin{array}{c}p \\ q\end{array}\right)=\frac{1}{2\Pi^2+\Pi(\log r^*)_T} \left(\begin{array}{cc}  v_T+v\Pi & v \\ 
u_T-u\Pi & u  \end{array}\right) \left(\begin{array}{c}\Pi_X\\ -\Pi_{XT}+2\Pi\end{array}\right), 
\eeq   
with  $\Pi=uv$ and $ r^*=\frac{v}{u}$. The involution (\ref{invol}) swaps $w\leftrightarrow\Pi$ and 
$r\leftrightarrow r^*$, and this leads to the alternative system (\ref{rstarfsys}), from which $u,v$ are recovered directly. 
\end{prf} 

\begin{figure}[ht]
\centering
\begin{subfigure}{.4\textwidth}
  \centering
  \includegraphics[width=\linewidth]{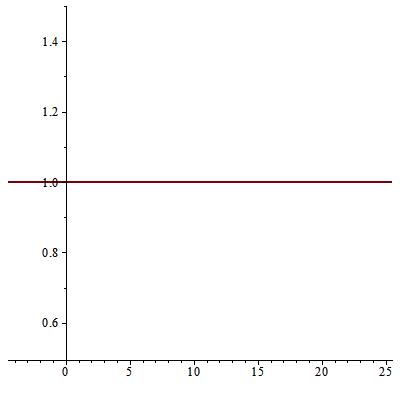}
\end{subfigure}%
\begin{subfigure}{.4\textwidth}
  \centering
\includegraphics[width=\linewidth]{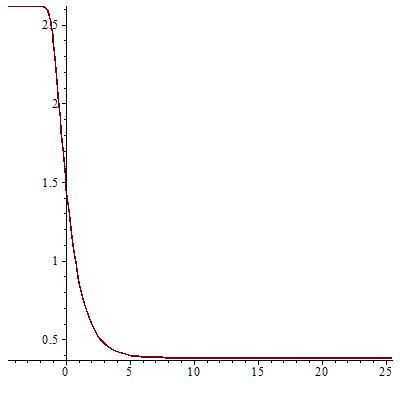}
\end{subfigure}
\caption{Parametric plots of $u$ and $v$ against $z$ for 
 (\ref{kinks}) with $c=-3$, $\rC =-1$, $A=1$.}
\label{kink}
\end{figure}

\begin{figure}
\centering
\begin{subfigure}{.5\textwidth}
  \centering
  \includegraphics[width=\linewidth]{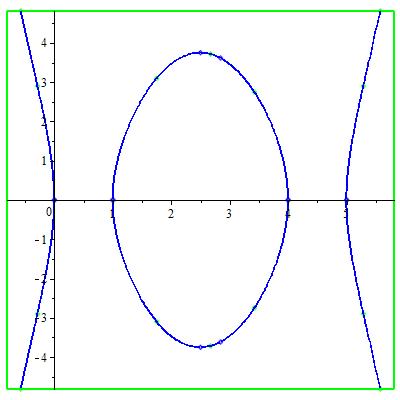}
  \caption{The quartic curve (\ref{piellip}) in the $(\pi,\pi ')$ plane.}
  \label{fig:sub1}
\end{subfigure}%
\begin{subfigure}{.5\textwidth}
  \centering
  \includegraphics[width=\linewidth]{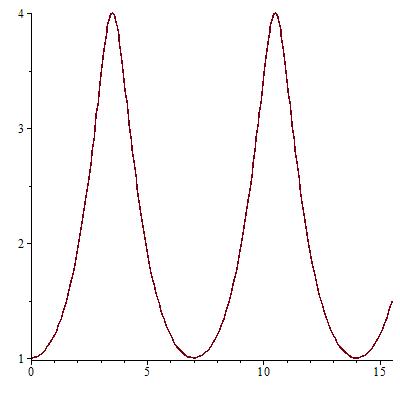}
  \caption{Parametric plot of $(z(Z),\pi(Z)$).} 
  \label{fig:sub2}
\end{subfigure}
\caption{Phase plane and $\pi$ against $z$ in the case $c=-5$, $\rC=-1$, $K=-1$. }
\label{fig:test}
\end{figure}

\noindent {\bf Example: travelling wave solutions.} To illustrate the preceding result, we consider travelling wave solutions 
of (\ref{cusys1}), such that $u$ and $v$ are functions of $z=x-ct$, where $c$ is the velocity of the waves. By comparing the conservation law (\ref{Fttau}), or the first equation in (\ref{rwsys}), with the reciprocal transformation    (\ref{curttau}) (where we ignore $\rd \tau$ and $\rd s$), it follows that such solutions correspond to travelling waves in the system 
(\ref{rtcub1}) which are functions of the variable $Z=X-\rC T$ for another constant $\rC$, where setting $pq=w(X,T)\to w(Z)$, $uv=\Pi(X,T)\to\pi(Z)$ yields 
\beq\label{wpi} 
\rC w=\pi +c.
\eeq 
Furthermore,  for the independent variables we have 
$$ 
\rd z = \rd x - c \rd t =w(X,T)\, \rd X - \Pi(X,T)\, \rd T - c\, \rd t =   w(Z)\, \rd X -(\pi(Z)+c)\rd T = w(Z)\,\rd Z, 
$$ 
by (\ref{wpi}). 
so if we replace $f(X,T)\to \tilde{f}(Z) +cT$ then 
\beq\label{zZ} 
z=\tilde{f}(Z), \qquad \text{with} \quad \tilde{f}'(Z)=w(Z)=\rC^{-1}(\pi(Z)+c), 
\eeq 
where the prime denotes $\rd /\rd Z$. 
To describe these travelling waves, it is most convenient to obtain a single equation for $\pi(Z)$, which is achieved by first using the definition of ${\cal B}^*$  to write 
\beq\label{rstar} 
(\log r^*)' = \rC^{-1}({\cal B}^*+2\pi), 
\eeq 
then putting this  and  (\ref{wpi}) into (\ref{rstarfsys}), to obtain a pair of quadratic equations in ${\cal B}^*$ with 
coefficients depending only on $\pi$ and its derivatives. After eliminating ${\cal B}^*$ 
to find  
\beq\label{bst} 
{\cal B}^*= -\frac{\rC \Big(2\rC \pi\pi ''-\rC(\pi ')^2+4\pi^2\Big)\Big(\rC\pi '' +4\pi +2c\Big)}{\pi (2\pi+c)\Big(\rC(\pi ')^2+4\pi^2+4c\pi\Big)}, 
\eeq 
then removing a prefactor, a single equation of second order and second degree for $\pi$ results: 
\beq\label{pieq} 
\Big(\pi '' + \rC^{-1}(4\pi+2c)\Big)^2 =\rC^{-2}\Big(4\pi^2+4c\pi+c^2\Big)(\pi ')^2+ \rC^{-3}\Big(16\pi^4+32c\pi^3+20c^2\pi^2+4c^3\pi\Big). 
\eeq  
The latter equation has a first integral: 
if $\pi$ satisfies the first order equation 
\beq\label{piellip} 
(\pi ')^2 =\frac{\pi^4}{\rC^{2}} +\frac{2c\pi^3}{\rC^{2}} +\left(\frac{c^2}{\rC^{2}}-4K\right)\pi^2 -4 Kc\pi +4(K\rC  -1)^2\equiv\mathrm{Q}(\pi ), 
\eeq 
for any constant value $K$, then it satisfies (\ref{pieq}). The generic solution of (\ref{piellip}) is an elliptic function of $Z$, 
but 
to have bounded periodic solutions 
for real $c,\rC,K$ requires 
that the curve $(\pi ')^2=\mathrm{Q}(\pi )$  
in the real $(\pi,\pi ')$ phase plane should have a compact component (see Figure \ref{fig:sub1}), 
otherwise solutions are generically unbounded with simple poles on the real $Z$ axis. 

The  quartic $\mathrm{Q}$ has  discriminant 
$\Delta = 256(2K\rC-1)^2 \rC^{-8}(16c^2\rC+(8K\rC^2+c^2-8\rC)^2)$.  
In order to obtain non-periodic bounded solutions, we fix $K=(2\rC)^{-1}$, so that $\Delta =0$ 
and (\ref{piellip}) gives 
\beq\label{sign}
\pi ' = \pm \rC^{-1}(\rC-c\pi-\pi^2). 
\eeq
Upon taking the plus sign above, this yields 
$$ 
\pi (Z)=\rC k\, \mathrm{tanh}(kZ)-\frac{c}{2}, \qquad k=\frac{\sqrt{c^2+4\rC}}{2\rC},  
$$ 
up to shifting the origin in $Z$, 
and then  using  (\ref{rstar}) and (\ref{bst}) we find 
$$ 
r^* (Z)=\exp \int  \rC^{-1}({\cal B}^*+2\pi)\rd Z =A\Pi(Z), 
$$ 
where $A>0$ is an arbitrary integration constant. 
Finally, from (\ref{zZ}) and Theorem \ref{uvparam}, we see that the solution of (\ref{cusys1}) is given parametrically by  
\beq\label{kinks}
z= \log\mathrm{cosh}(kZ)+\frac{cZ}{2\rC}, \qquad u=\frac{1}{\sqrt{A}}, 
\qquad v=\sqrt{A}\left(\rC k\,\mathrm{tanh}(kZ)-\frac{c}{2}\right), 
\eeq 
up to shifting $z$ by an arbitrary constant. 
It is necessary to impose the conditions 
$$ 
c<0,\qquad \rC<0, \qquad c^2+4\rC>0, 
$$ 
in order to have a real single-valued solution in $z$,  otherwise $\frac{\rd z}{\rd Z}=w(Z)$ will vanish for some $Z$. 
So in this solution, corresponding to the plus sign in (\ref{sign}), $u$ is constant and $v$ is a kink-shaped travelling wave - 
see Figure \ref{kink}; 
with the opposite choice of sign, the roles of $u$ and $v$ are reversed.   

\begin{figure}
\centering
\begin{subfigure}{.5\textwidth}
  \centering
  \includegraphics[width=\linewidth]{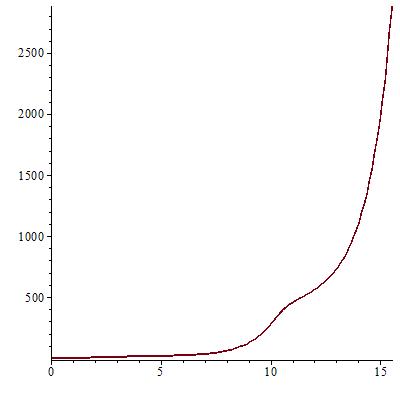}
  \caption{Parametric plot of $(z(Z),u(Z)$).}
  \label{fig:u}
\end{subfigure}%
\begin{subfigure}{.5\textwidth}
  \centering
  \includegraphics[width=\linewidth]{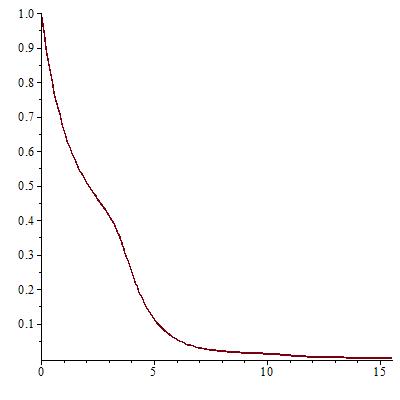}
  \caption{Parametric plot of $(z(Z),v(Z)$).} 
  \label{fig:v}
\end{subfigure}
\caption{Travelling wave profiles of  $u$ and $v$ against $z$ with $c=-5$, $\rC=-1$, $K=-1$, $A=1$. }
\label{fig:uv}
\end{figure}
To obtain explicit formulae for travelling waves in general, one should fix a root $\pi_0$ of the quartic $\text{Q}$ in 
(\ref{piellip}), and make a birational change of variables of the form $\wp = \al (\pi- \pi_0)^{-1}+\be $, 
$\wp'=-\al \pi ' (\pi- \pi_0)^{-2}$, to yield a cubic equation of the form $(\wp ')^2=4\wp^3-g_2\wp -g_3$ for the 
Weierstrass $\wp$-function. For example, the special case $c=-m^2-1$, $\rC=-1$, $K=-m^2/2-m-1$ gives a one-parameter 
family of quartics, which has the root $\pi_0=-1$  
for all values of the parameter $m$, and has 4 real roots whenever $m<-1$ or $m>3$, giving a curve with a compact oval 
(as in Figure \ref{fig:sub1}). To illustrate the form of the solution, we fix $m=-2$, so that 
Q$=\pi (\pi-1)(\pi-4)(\pi-5)$, and find 
\beq\label{pipfn} 
\pi (Z) = \frac{3}{ \wp (Z)-\frac{5}{12} } + 1 =  \frac{\wp '(Z^*)}{\wp (Z)-\wp(Z^*)} + 1, 
\eeq 
where $\wp (Z)$ denotes the $\wp$-function with invariants $g_2=\frac{241}{12}$, $g_3=-\frac{3689}{216}$ and 
half-periods $\om_1\approx 1.40060304$, $\om_2\approx .79812111 i$, and 
$$ 
 Z^*=-\int_{5/12}^{7/6} 
\frac{\rd \zeta} { \sqrt{4\zeta^3-\frac{241}{12}\zeta +\frac{3689}{216}} }
+\om_1+\om_2\approx .70030152+.79812111 i. 
$$ 
Using (\ref{wpi}), together with (\ref{rstar}) and (\ref{bst}), we see that 
$$ 
(\log r^*)' = -w =  \frac{\wp '(Z^*)}{\wp (Z)-\wp(Z^*)} - 4, 
$$ 
and upon integration this yields 
\beq \label{rexp} 
r^*(Z)=A\, \frac{\si (Z^*-Z)}{\si (Z^*+Z)}\, \exp (2\zeta(Z^*)z-4z), \qquad  
z= -\log r^*(Z),  
\eeq 
up to shifting $z\to z+$const, where the constant $A>0$ is arbitrary. As a function of $Z$, the product $\pi = uv$ given 
by (\ref{pipfn}) has real period $2\om_1$, and from (\ref{rexp}) it follows that it is also periodic in $z$: 
when $Z\to Z+2\om_1$ then $z\to z+\Omega$, 
where $\Omega =-\log |\exp (4\zeta (Z^*)\om_1-4\zeta (\om_1)Z^*) |+8\om_1\approx 7.00301521$ is the 
period (see Figure \ref{fig:sub2}). On the other hand, from (\ref{rexp}) we also have $r^*=\exp(-z)$, so by 
Theorem \ref{uvparam} the travelling wave profiles of $u=\sqrt{\pi/r^*}$ and $v=\sqrt{\pi r^*}$ consist of 
 exponentially/growing decaying solutions on a periodic background  (see Figure \ref{fig:uv}).

\subsection{Second cubic system} 

After removing the linear dispersion terms and rescaling for the sake of simplicity, the system (\ref{sysc2}) 
becomes 
\beq\label{cusys2} 
m_t=(uvm)_x, \qquad n_t = (uvn)_x, 
\qquad 
\text{with} \quad m=u-u_x, \, n= v+v_x .
\eeq
Both equations in this system are in conservation form, but in order to apply 
a reciprocal transformation we pick the conservation law 
\beq\label{conq} 
q_t = (p\,q)_x, \qquad \text{where} \quad q=(m n)^{\frac{1}{2}}, \quad p=uv.
\eeq 
For what follows, we also note the equation 
\beq\label{ceq} 
\ka_t = p\,\ka_x, \qquad \text{where} \quad \ka=(n /m)^{\frac{1}{2}}. 
\eeq 
Now from (\ref{conq}) we can define new independent variables according to 
\beq \label{recipc} 
\rd X=q\, \rd x +pq\,\rd t, \qquad \rd T = \rd t, 
\eeq 
so that derivatives transform according to $\partial_x = q\,\partial_X$, 
$\partial_t = \partial_T +pq\,\partial_X$. Since this is a reciprocal transformation, the equation 
(\ref{conq}) becomes a conservation law in the new variables, that is 
\beq\label{qT} 
\left(\frac{1}{q}\right)_T +p_X=0, 
\eeq 
while the evolution of $\ka$ in (\ref{ceq}) becomes 
$$
\frac{\partial \ka}{\partial T}=0\implies \ka=\ka(X).  
$$
This means we can write the quantities 
$m$ and $n$ 
in terms of $q$ as  
\beq\label{exi} 
m = \ka^{-1}\,  q, \qquad n =\ka\, q, 
\eeq  
where the prefactor $\ka^{\mp 1}$ depends only on the new independent variable $X$. 
The question is now how to find an equation for $q=q(X,T)$ and thence obtain the fields $u$ and $v$ in terms of functions of $X$ and $T$, and thence obtain 
solutions $u(x,t)$, $v(x,t)$ in parametric form. 

To begin with note that, in view of (\ref{recipc}) and (\ref{exi}), we can use 
$u_x=u-m$, $v_x=n-v$  
and transform the derivatives to find  
\beq\label{uvx} 
u_X=q^{-1}\, u-\ka^{-1}, \qquad v_X =\ka-q^{-1}\, v.
\eeq 
This means that from (\ref{qT}) we obtain 
$
\partial_T (q^{-1}) =-(u_Xv+uv_X)= 
-\ka\, u+\ka^{-1}v 
$, 
and hence 
\beq\label{vsub} 
v = \ka^2\, u+\ka\, (q^{-1})_T. 
\eeq 
The above expression for $v$ can be substituted back into  (\ref{qT}) to yield 
\beq\label{evolq} 
\left(\dfrac{1}{q}\right)_T =-\frac{\partial}{\partial X} \Big( u\, (\ka\, (q^{-1})_T-\ka^2\, u)\Big). 
\eeq 
In order to get a single equation involving only $\ka$ and $q$, it is necessary to write $u$ in terms of 
$\ka$,$q$ and their derivatives, and this is achieved by substituting (\ref{vsub}) into the second 
equation in (\ref{uvx}), so that the latter becomes a linear system for $u$ and $u_X$, which is readily solved. 
However, 
it turns out that it is most convenient to introduce a new function $\ups 
(X,T)$, which is defined by 
\beq\label{updef} 
\frac{1}{q} = 2\ups - \frac{\rd}{\rd X} \log \ka(X). 
\eeq 
In terms of $\ups$ and $\ka$, $u$ and $v$ are then given by 
\beq\label{uvform} 
u = \ka^{-1}\, \frac{(1-\ups_{XT}-2\ups\ups_T)}{2\ups} , 
\quad 
v = \ka\, \frac{(1-\ups_{XT}+2\ups\ups_T)}{2\ups},  
\eeq 
so that the product $p=uv$ is independent of $\ka$, and so  (\ref{qT}), or equivalently (\ref{evolq}), becomes 
an autonomous partial differential equation for $\ups$ alone, namely 
\beq\label{upeq} 
\ups_T = \frac{1}{2}\frac{\partial}{\partial X}\left(\ups_T^2-\frac{(\ups_{XT}-1)^2}{4\ups^2} \right). 
\eeq 
Upon introducing a potential $f(X,T)$  such that $\ups = f_X$, this equation can be integrated with respect to $X$, and an arbitrary function of $T$ that appears can be absorbed into $f$ without loss of generality, so that an equation of third order for $f$ results, that is 
\beq\label{feq} 
(f_{XXT}-1)^2 -4f_X^2f_{XT}^2+8f_X^2f_T=0. 
\eeq 
\begin{The} \label{xch4} 
Let $f=f(X,T)$ be a solution of (\ref{feq}), let $\ka=\ka(X)$ be an arbitrary function, and let 
$\ups (X,T)=f_X(X,T)$.
Then setting $$x=2f(X,t)-\log \ka(X)$$ together with (\ref{uvform}) gives a  solution $(u(x,t),v(x,t))$
of the system (\ref{cusys2})
in parametric form. 
\end{The} 
\begin{prf} 
Comparison of (\ref{feq}) with (\ref{uvform}) shows that $p=uv=-2f_T$. Then 
taking the differential of $x$ above gives $\rd x =
 (2f_X(X,t)-\partial_X \log \ka)\, \rd X+2f_T(X,t) \, \rd t = q^{-1} \rd X - p\,\rd T$,  
in accordance with the inverse of the reciprocal transformation (\ref{recipc}). By reversing the reciprocal transformation, the 
equations (\ref{conq}) and (\ref{ceq}) result, and together these imply the system (\ref{cusys2}) for $u$ and $v$.
\end{prf}
In order to find solutions of the equation (\ref{feq}), it is instructive to consider the behaviour near singularities. The equation has two types of expansions near a movable singularity manifold $\varphi(X,T)=0$, with leading order behaviour 
$f\sim \pm\log\varphi$, corresponding to simple poles in the solution of (\ref{upeq}). This suggests that one can apply the 
two-singular-manifold method introduced in \cite{CM}, leading to the following result. 
\begin{Pro} 
The equation (\ref{feq}) has an auto-B\"{a}cklund transformation which relates two solutions $f,\hat f$ according to the 
transformation 
$$ 
\hat{f}=\log Y+f, 
$$ 
where $Y$ is a solution of the Riccati system 
\beq\label{ric} \bear{rcl} 
Y_X & = & \lambda -2f_X\, Y +Y^2, \\
Y_T & = & \left(\dfrac{f_{XT}}{2}+\dfrac{(1-f_{XXT})}{4f_X}\right)  -\dfrac{1}{2\lambda}\, Y + 
\left(-\dfrac{f_{XT}}{2}+\dfrac{(1-f_{XXT})}{4f_X}\right)\, Y^2 
\eear 
\eeq 
with an arbitrary parameter $\la$. The Riccati system is linearized via the transformation 
$$Y^{-1}=\frac{1}{\la}\left(\frac{\psi_X}{\psi}+\ups\right),$$ to yield a scalar 
Lax pair for (\ref{upeq}), given by 
\beq\label{laxpsi}\bear{l} 
\psi_{XX}+(\ups_X-\ups^2+\la)\psi = 0, \\ 
\psi_T = \la^{-1}({\cal U}\psi_X-\frac{1}{2}{\cal U}_X \psi), \qquad 
{\cal U} = \dfrac{\ups_T}{2}+\dfrac{(1-\ups_{XT})}{4\ups_X}. 
\eear 
\eeq  
\end{Pro}
\begin{Cor} 
The system (\ref{cusys2}) has the scalar  Lax pair 
\beq\label{laxphi}\bear{l} 
\phi_{xx}+(q^2\la +r)\phi = 0, \\ 
\phi_t = (p+w\la^{-1})\phi_x-\frac{1}{2}(p_x+w_x\la^{-1})\phi, \qquad 
\eear 
\eeq  
where 
\beq\label{rwdef} 
r=-\frac{w_{xx}}{2w}+\frac{w_{x}^2}{4w}-\frac{1}{16w^2}, \quad 
w = \frac{1}{4q}\left( \frac{\Big((p_x \, q^{-1})_x+2q\Big)}{\Big(1+(\log \ka )_x\Big)}- \frac{p_x}{ q}\right), 
\eeq 
with $p=uv$, 
$q=\sqrt{mn}$, $\ka=\sqrt{n/m}$ 
as above.  
\end{Cor} 
{\bf Proof of Corollary:} 
The Lax pair follows from (\ref{laxpsi}) by setting $\psi = \sqrt{q}\,\phi$ and applying the inverse 
of the reciprocal transformation (\ref{recipc}). The compatibility conditions for this linear system 
consist of  (\ref{conq}) together with 
$$ 
r_t=\frac{1}{2}p_{xxx} +2p_x r +pr_x +2q^2 w_x +2qq_xw \quad \text{and} \quad 
w_{xxx}+4rw_x+2r_xw=0,
$$ 
where the last one is a consequence of the definition of $r$ in (\ref{rwdef}). These conditions 
are best checked with computer algebra. \qed 
  
The form of the Lax pair (\ref{laxpsi}) reveals that $\ups$ corresponds to the dependent variable for the modified KdV equation, and the standard Miura map ${\cal V}= \ups_X - \ups^2$ relates (\ref{upeq}) to the first negative flow of the KdV 
hierarchy, as considered in   \cite{Fuchss} (see also \cite{WH}), which takes the form 
$$ 
{\cal V}_T = 2{\cal U}_X, \qquad 
{\cal U}{\cal U}_{XX} -\frac{1}{2}{\cal U}_X^2 +2{\cal V}{\cal U}^2 + \frac{1}{8}=0 
$$ 
in terms of the variables ${\cal U}$, ${\cal V}$. If $r$ and $w$ were constants, then (\ref{laxphi}) would reduce to 
the Lax pair for the Camassa-Holm equation, as presented in \cite{CH}. 

\begin{figure}
\centering
\begin{subfigure}{.5\textwidth}
  \centering
  \includegraphics[width=0.8\linewidth]{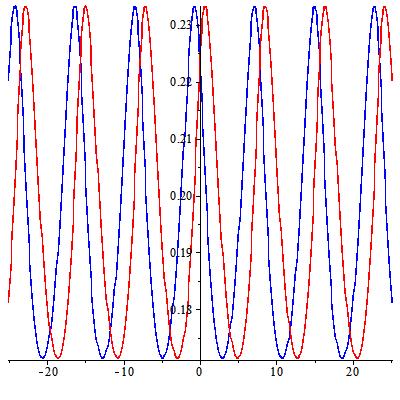}
  \caption{Plots of $u$ and $v$ at $t=0$ with $\kappa=1$.}
  \label{fig:uv1}
\end{subfigure}%
\begin{subfigure}{.5\textwidth}
  \centering
  \includegraphics[width=0.8\linewidth]{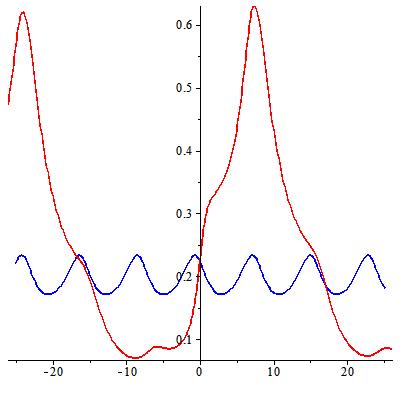}
  \caption{ Plots of $u$ and $v$ at $t=0$ with $\kappa=\exp(\sin X)$.}
  \label{fig:uvs}
\end{subfigure}
\caption{Parametric plots of  $u$ (blue) and $v$ (red) against $x$ for different choices of $\ka (X)$. }
\label{fig:uvch4}
\end{figure}
\noindent
{\bf Example: periodic solutions and their deformations.} 
To obtain simple solutions of the system (\ref{cusys2}), we consider solutions of (\ref{feq}) which, apart from a shift 
by a linear function of $T$, depend only on the travelling wave variable $Z=X-\mu T$. Upon setting  
$$ 
f(X,T)=\tilde{f}(Z)-\nu T, 
$$ 
we find that $W(Z)=\tilde{f}'(Z)$ satisfies the following ordinary differential equation of second order and second degree: 
\beq\label{w2o2d} 
(\mu W''+1)^2 
-4\mu^2  W^2 (W')^2 - 8(\mu W+\nu)W^2 =0. 
\eeq 
The latter equation is solved in elliptic functions: for any value of the constant $c_2$, $W$ is a solution of (\ref{w2o2d}) 
whenever it satisfies 
\beq\label{wquart} 
(W')^2=W^4 +c_2 W^2 -2\mu^{-1} W +c_0, \qquad c_0=\frac{c_2^2}{4}-\frac{2\nu}{\mu^2}.
\eeq 
For such a solution, Theorem \ref{xch4} gives 
\beq\label{xeq} 
x= \log \left(\frac{ \rho (X-\mu t )^2}{\ka (X)}
\right) -\nu t, 
\qquad \text{with} \quad \rho (Z)=\exp \int W(Z)\,\rd Z, 
\eeq 
while (\ref{uvform}) becomes 
\beq\label{uvwform} 
u = \ka^{-1}\, \frac{(1+\mu W''+2\mu WW')}{2W} , 
\quad 
v = \ka\, \frac{(1+\mu W''-2\mu WW')}{2W},  
\eeq 
so in order to avoid singularities in $u$ and $v$, we require that $W$ should be a bounded, positive periodic function of $Z$; 
this is achieved by choosing the quartic on the right-hand side of (\ref{wquart}) to have three positive real roots, 
$0<w_1<w_2<w_3$, whence the fourth root is $w_0=-(w_1+w_2+w_3)<0$. Using a M\"{o}bius transformation 
$W=\al (\wp -\be )^{-1} +w_1$ to send the first positive root to infinity leads to the solution in terms of Weierstrass functions, similarly to the previous example for the system (\ref{cusys1}). 

For illustration, we pick the quartic $(W+9)(W-2)(W-3)(W-4)$ in (\ref{wquart}), so that 
$c_2=-55,c_0=-216, \mu = -1/105,\nu=3889/88200$, and then 
\beq\label{wp} 
W(Z) =  \frac{11}{2\wp(Z)+\frac{31}{6}} + 2 = \frac{\wp ' (Z^*)}{\wp (Z) - \wp (Z^*)} +2, 
\eeq  
where the $\wp$ function is associated with the cubic $(\wp ')^2=4\wp^3-\frac{433}{12}\wp +\frac{1295}{216}$
with half-periods $\om_1\approx 0.77203133$, $\om_2\approx.74313318 i$, and 
$$ 
 Z^*=-\int_{-31/12}^{1/6} 
\frac{\rd \zeta} { \sqrt{4\zeta^3-\frac{433}{12}\zeta +\frac{1295}{216}} }
+\om_1+\om_2\approx .16697654+.74313318 i.
$$ 
For the function $\rho$ in (\ref{xeq}) we find  
$$ 
\rho (Z) = \frac{\si (Z^*-Z)}{\si (Z^* + Z)} \, \exp (2\zeta (Z^*) Z + 2Z). 
$$ 
The behaviour of the solutions $u(x,t)$, $v(x,t)$ obtained in this way depends crucially on the choice of function 
$\ka (X)$. 
In order to have singled-valued soutions 
it is necessary that the derivative $\partial x/\partial X$ should never vanish, which requires that the logarithmic 
derivative $\ka ' /\ka$ should be suitably bounded. In particular, if $\ka=\,$constant then this is so, and in that case 
travelling wave solutions of (\ref{cusys2}) result, and both $u$ and $v$ are periodic functions. More generally, taking 
$\ka = \exp k(X)$ in (\ref{uvwform}), where both the function $k$ and its first derivative are bounded, gives bounded deformations of these periodic solutions - see Figure \ref{fig:uvch4} for the comparison between the cases 
$\ka=1$ and $\ka=\exp\sin X$. However, if $\ka = \exp k(X)$ with $k(X)$ being a linear function of $X$, then 
unbounded solutions result, exhibiting similar profiles to the solutions of (\ref{cusys1}) with  exponential growth/decay on a periodic background, as illustrated in Figure \ref{fig:uv}.

\section{Conclusions} 

The perturbative symmetry approach has yielded a classification of integrable two-component systems of the form 
(\ref{CHsys}), producing two systems with quadratic nonlinearities (Theorem 2), two systems with cubic nonlinearities  (Theorem 3), 
and two mixed quadratic/cubic systems  (Theorem 4); the systems with mixed nonlinear terms include the others as limiting cases, by sending suitable parameters to zero. At the same time, an alternative approach via compatible Hamiltonian operators 
has provided a different set of  two-component systems, and has allowed us to find bi-Hamiltonian structures for all of the systems obtained from the symmetry approach. We have also found Lax pairs for all of the systems in Theorems 2 and 3, at least in the absence of linear dispersion terms, as well as reciprocal transformations linking them to known integrable hierarchies, and this has allowed us to construct some simple solutions explicitly. 

As far as we know, integrable systems of the form (\ref{CHsys}) have not been considered in detail before, 
apart from Falqui's system (\ref{falqui}). However, while we were completing this work 
we learned of a three-component system in  
which two of the equations 
involve nonlocal terms of  this type; the system was constructed as a dispersive version of the WDVV 
associativity equations \cite{pavlov}. There are several issues still to be resolved regarding the systems  
introduced here. In particular, for the systems (\ref{sysq2}), (\ref{sysc1}) and (\ref{sysc2}), as well as the systems 
in Theorem 4, we have not presented Lax pairs that include the linear dispersion terms. Also, the system (\ref{eqh1}), 
or equivalently 
 (\ref{eqh2}), is worthy of further analysis, since it is outside the class  (\ref{CHsys}). 

In the near future, we further intend to classify  two-component systems  with the nonlocal terms 
$(1-D_x^2)u_t$, $(1-D_x^2)v_t$ on the left-hand side, which include (\ref{eqh6}). Recently, various 
different systems of this kind 
have been proposed \cite{song, xiaq}, which deserve to be studied further. 

\noindent 
{\bf Acknowledgments:} ANWH is supported by Fellowship EP/M004333/1 from the Engineering and Physical Sciences 
Research Council (EPSRC). 
JPW and VN were partially supported by Research in Pairs grant no. 41418 from the 
London Mathematical Society; 
VN also thanks the University of Kent for the hospitality received during his visit in June 2015,
funded by the grant. In addition, JPW was supported by the EPSRC grant EP/1038659/1.

\end{document}